\documentclass[final,3p,times,authoryear]{elsarticle}

\usepackage{amssymb}

\usepackage{lineno}

\journal{Ocean Modelling}

\usepackage{color}
\usepackage{amsmath}

\newcommand{\dy}{\partial}
\newcommand\ddy[2]{\frac{\dy#1}{\dy#2}}

\newcommand{\grad}{\nabla}

\newcommand{\nb}{\boldsymbol{n}}

\newcommand{\ub}{\boldsymbol{u}}

\newcommand{\rhobar}{{\overline{\rho}}}

\definecolor{dark-green}{rgb}{0,0.5,0} 

\usepackage{rotating}
\usepackage{relsize}

\begin{document}

\begin{frontmatter}



\title{Emergent eddy saturation from an energy constrained eddy parameterisation}


\author[edinburgh]{J. Mak\fnref{tel}}

\fntext[tel]{Corresponding author. Tel: +44 131 650 5040}
\ead{julian.c.l.mak@googlemail.com}

\author[oxford]{D. P. Marshall}

\author[edinburgh]{J. R. Maddison}

\author[cambridge]{S. D. Bachman}

\address[edinburgh]{School of Mathematics and Maxwell Institute for Mathematical
Sciences, University of Edinburgh, Edinburgh, EH9 3FD, United Kingdom}

\address[oxford]{Department of Physics, University of Oxford, Oxford, OX1 3PU,
United Kingdom}

\address[cambridge]{DAMTP, Centre for Mathematical Sciences, Wilberforce Road,
Cambridge, CB3 0WA, United Kingdom}



\begin{abstract}
The large-scale features of the global ocean circulation and the sensitivity of
these features with respect to forcing changes are critically dependent upon the
influence of the mesoscale eddy field. One such feature, observed in numerical
simulations whereby the mesoscale eddy field is at least partially resolved, is
the phenomenon of eddy saturation, where the time-mean circumpolar transport of
the Antarctic Circumpolar Current displays relative insensitivity to wind
forcing changes. Coarse-resolution models employing the Gent--McWilliams
parameterisation with a constant Gent--McWilliams coefficient seem unable to
reproduce this phenomenon. In this article, an idealised model for a
wind-forced, zonally symmetric flow in a channel is used to investigate the
sensitivity of the circumpolar transport to changes in wind forcing under
different eddy closures. It is shown that, when coupled to a simple
parameterised eddy energy budget, the Gent--McWilliams coefficient of the form
described in \cite{Marshall-et-al12} [\emph{A framework for
parameterizing eddy potential vorticity fluxes}, J. Phys. Oceanogr., vol. 42,
539--557], which includes a linear eddy energy dependence, produces eddy
saturation as an emergent property.
\end{abstract}


\end{frontmatter}



\section{Introduction}

Studies of the response of the large-scale ocean circulation to changing forcing
scenarios in numerical ocean models, and its resulting climatology, require long
time integrations that are prohibitively expensive even at mesoscale eddy
permitting resolutions. Since this is expected to remain the case for the
foreseeable future, an ongoing challenge in numerical ocean modelling is the
representation of the unresolved mesoscale eddy field in coarse resolution
models. A particularly successful scheme that is employed is the
Gent--McWilliams closure \citep[][hereafter GM]{GentMcWilliams90, Gent-et-al95},
which parameterises mesoscale eddies via the introduction of a non-divergent
eddy transport velocity. The eddy transport velocity can be interpreted as
arising from the difference between the Eulerian average of the velocity at
fixed height and the thickness-weighted average of the velocity at fixed density
\citep{McDougallMcIntosh01}, and modifies the advective transport of tracer
quantities. By definition, the non-divergent eddy transport velocity conserves
all moments of the advected quantities, and is thereby adiabatic. The property
of adiabatic stirring is particularly attractive, being shown to remove spurious
heating and cooling in the deep ocean, such as that associated with the Deacon
cell in the Southern Ocean \citep{Danabasoglu-et-al94}.

To this point, studying the modelled oceanic response to changing atmospheric
forcing in conjunction with the GM parameterisation is of particular importance
for emergent climatologies under different forcing scenarios. Two important
large-scale Southern Ocean phenomena are of particular interest in this regard.
The first is ``eddy saturation'', originally discussed in \cite{Straub93} from
an argument based on critical stability, and here to be understood as the
relative insensitivity of the time-mean circumpolar transport with respect to
wind forcing changes. The other is ``eddy compensation'', here to be understood
as the reduced sensitivity of the residual meridional overturning circulation
with wind forcing changes \citep[e.g.,][]{Meredith-et-al12, ViebahnEden12,
Munday-et-al13}, which has consequences for the meridional transport of
important tracers such as heat, salt and carbon. This article focuses on eddy
saturation.

As argued by \cite{Straub93}, if fluid interaction with topography is the main
sink for momentum input by wind stress, and consequently the zonal abyssal flow
is weak, then thermal wind shear is the dominant contribution to circumpolar
transport. Thus circumpolar transport is intimately linked to isopycnal slope,
with the slope steepness limited by baroclinic instability. Eddy saturation
arises through a balance between steepening of isopycnals by wind stress, and
flattening of isopycnals by the presence of the mesoscale eddy field. The
reduction in circumpolar transport sensitivity with varying wind stress has been
observed in a variety of numerical models that at least partially resolves a
mesoscale eddy field \cite[e.g.,][]{HallbergGnanadesikan06, HoggBlundell06,
Hogg-et-al08, FarnetiDelworth10}. In \cite{Munday-et-al13}, an eddy permitting
one-sixth degree model of a 20 degree wide ocean sector was integrated with
varying wind forcings. This eddy permitting model, employing a very small value
of the GM coefficient, showed near complete eddy saturation. By contrast, in
lower resolution half degree and two degree variants of the same model, where
larger values of the GM coefficient were utilised, the resulting time-mean
circumpolar transport displayed significant sensitivity with respect to the wind
forcing. \cite{HoggMunday14} found that although the value of the time-mean
circumpolar transport was affected by the domain geometry, the relative
insensitivity with changing wind stress at eddy permitting resolution was
robust.

Thus it has been found that the GM scheme with spatially and temporally constant
GM coefficient is unable to represent eddy saturation. With increased wind
forcing, a more vigourous eddy field is to be expected. Since the GM coefficient
in some sense specifies the intensity and efficiency of the parameterised eddy
field, it is expected that a positive correlation between the strength of wind
forcing and the magnitude GM coefficient is minimally required for emergent eddy
saturation. Various proposals already exist with a non-constant GM coefficient.
\cite{Visbeck-et-al97}, using linear stability arguments, proposes a GM
coefficient which depends upon the stratification profile, as well as a mixing
length. In \cite{Ferreira-et-al05} the eddy-mean-flow interaction in a global
ocean model is determined via an optimisation procedure, yielding diagnosed
values for the GM coefficient. This is used to infer a GM coefficient which
depends on the vertical stratification, and has subsequently been incorporated
into a number of ocean general circulation models
\citep[e.g.,][]{DanabasogluMarshall07, GentDanabasoglu11}. The simulations
described in \cite{GentDanabasoglu11} do show some eddy compensation, as a
consequence of the dependence of the GM coefficient on Southern Ocean
stratification. However, as discussed in \cite{Munday-et-al13}, this mechanism
precludes the model from achieving full eddy saturation. A case where the GM
coefficient is varied manually with changing wind stress has also been
investigated \citep{Fyfe-et-al07}. Through the consideration of the eddy kinetic
energy budget, \cite{Cessi08} proposes a mixing length based eddy
parameterisation, with a GM coefficient depending on the ocean state and also
explicitly depending on the strength of the bottom drag. An approach also based
upon consideration of the eddy kinetic energy budget is discussed in
\cite{EdenGreatbatch08}, also employing a mixing length argument but utilising a
local parameterised eddy kinetic energy budget to inform the magnitude and
spatial structure of the resulting GM coefficient.

In \cite{Marshall-et-al12} a geometric interpretation of the eddy-mean-flow
interaction for the quasi-geostrophic equations was derived. A horizontally
down-gradient closure for the horizontal eddy buoyancy fluxes leads to a GM
coefficient of the form
\begin{linenomath*}\begin{equation}\label{eqn:mmb_intro}
  \kappa = \alpha E \frac{N}{M^2},
\end{equation}\end{linenomath*}
where $E$ is the total (kinetic plus potential) eddy energy, and $N / M^2 = T$
is an Eady time-scale which depends on the mean stratification, with $N^2 =
-(g/\rho_0)\dy \overline{\rho} / \dy z$ and $M^2 = (g/\rho_0)\left| \nabla_H
\overline{\rho} \right|$, where $g$ is the gravitational acceleration, $\rho_0$
is a reference density, $\overline{\rho}$ is the mean density, and $\nabla_H
\overline{\rho}$ is its horizontal gradient. A crucial point is that, if the
eddy energy is known, there are no undetermined dimensional parameters; the only
freedom is to specify the non-dimensional geometric parameter $\alpha$ of
magnitude less than or equal to one.

This article assesses the ability of the \cite{Marshall-et-al12} for of the GM
coefficient in producing emergent eddy saturation, via numerical calculations in
an idealised, zonally averaged, two-dimensional ocean channel model. The
idealised numerical model is motivated by the physical model discussed in
\cite{Marshall-et-al16}. The performance is compared against a number of
alternative approaches, including approaches based upon mixing length arguments,
and based upon the \cite{Visbeck-et-al97} proposal. Since the
\cite{Marshall-et-al12} variant requires information about the eddy energy, the
evolution of the mean state is coupled to a simple prognostic equation for the
parameterised domain integrated eddy energy (cf. the local budget for the eddy
kinetic energy in \citealt{EdenGreatbatch08}).

A form similar to \eqref{eqn:mmb_intro} also appears in \cite{Jansen-et-al15}
--- implied by their equations (9) and (11) --- but with the eddy kinetic energy
in place of the full eddy energy, and motivated by the inverse energy cascade
being controlled by the rate of eddy energy generation through baroclinic
instability as per \citet{LarichevHeld95}. However the form derived in
\cite{Marshall-et-al12} provides an explicit upper bound on the relevant
geometric parameter $\alpha$. Hence no other dimensional scaling is possible
provided the geometric parameter $\alpha$ is bounded away from zero. Moreover
here the eddy energy is determined prognostically via the solution of a
dynamical equation which is coupled to the equations for the mean state.

The paper proceeds as follows. In \S2 the GM scheme and the
\cite{Marshall-et-al12} parameterisation variant are revisited, focusing in
particular on the energetics of the problem, and providing physical and
mathematical arguments as to why the \cite{Marshall-et-al12} variant may be
expected to have skill in producing emergent eddy saturation. \S3 contains the
details of the idealised numerical model and details of the other
parameterisation variants considered in this work. The actual implementation of
the parameterisation variants and their results are presented in \S4 for a case
where the GM coefficient is assumed to be constant over the domain, and in \S5
for a case where the GM coefficient is spatially varying, focusing on the case
where a spatial structure depending upon the vertical stratification is
enforced. The paper concludes in \S6, where the results are discussed, and a
recipe for implementation in a general global circulation models is proposed.


\section{Gent--McWilliams and energetic constraints}\label{sec:param}


\subsection{The Gent--McWilliams scheme and the energetic consequences}\label{sec:gm}

The GM scheme parameterises the effects of baroclinic eddies via the
introduction of an adiabatic stirring of the mean density, acting to decrease
the available potential energy of the system \citep[e.g.,][]{GentMcWilliams90}.
Limiting consideration to the Boussinesq case, the mean density equation,
zonally averaged at constant density \citep{Andrews83, McDougallMcIntosh01,
Young12}, is
\begin{linenomath*}\begin{equation}\label{eqn:mean_density}
  \ddy{\rhobar^\rho}{t} + \ddy{\left( \hat{v} ~ \rhobar^\rho \right)}{y} + \ddy{\left( \hat{w} ~ \rhobar^\rho \right)}{z}  = 0,
\end{equation}\end{linenomath*}
where $\overline{\left(\ldots\right)}^\rho$ indicates a zonal average at constant
density, with
\begin{linenomath*}\begin{equation}
  \hat{v} = \overline{v \left( \ddy{\rho}{z} \right)^{-1}}^\rho \ddy{\rhobar^\rho}{z}
\end{equation}\end{linenomath*}
the thickness-weight averaged meridional velocity at constant density, and
$\hat{w}$ defined such that
\begin{linenomath*}\begin{equation}
  \ddy{\hat{v}}{y} + \ddy{\hat{w}}{z} = 0.
\end{equation}\end{linenomath*}
Following \cite{McDougallMcIntosh01},
\begin{linenomath*}\begin{equation}
  \hat{\ub} = \left( \begin{array}{c} 0 \\ \hat{v} \\ \hat{w} \end{array} \right)
            = \left( \begin{array}{c} 0 \\ \overline{v}^z \\ \overline{w}^z \end{array} \right)
            + \left( \begin{array}{c} 0 \\ v^* \\ w^* \end{array} \right)
            = \overline{\ub}^z + \ub^*,
\end{equation}\end{linenomath*}
where $\overline{\ub}^z$ is the velocity zonally averaged at constant height,
and $\ub^*$ is the eddy transport velocity, with
\begin{linenomath*}\begin{equation}
  \ddy{\overline{v}^z}{y} + \ddy{\overline{w}^z}{z} = \ddy{v^*}{y} + \ddy{w^*}{z} = 0.
\end{equation}\end{linenomath*}
The GM scheme then takes the form
\begin{linenomath*}\begin{equation}\label{eq:u_eddy}
  \ub^* = \grad \times \left( \begin{array}{c} -\kappa s \\ 0 \\ 0 \end{array} \right)
        = \left( \begin{array}{c} 0 \\ -\dy \left( \kappa s \right) / \dy z \\ \dy \left( \kappa s \right) / \dy y \end{array} \right),
\end{equation}\end{linenomath*}
where $\kappa$ is the GM eddy transfer coefficient, and $s$ is the slope of the
mean density surfaces
\begin{linenomath*}\begin{equation}
  s = - \left( \frac{\partial \overline{\rho}^\rho}{\partial y} \right) \left( \frac{\partial \overline{\rho}^\rho}{\partial z} \right)^{-1}.
\end{equation}\end{linenomath*}

The energetic consequences of the GM scheme are as follows. Consider the zonally
averaged hydrostatic Boussinesq equations in the form
\begin{linenomath*}\begin{subequations}\label{eqn:mean_momentum}
  \begin{align}
    \ddy{\overline{u}^z}{t} + \overline{v}^z \ddy{\overline{u}^z}{y}
                            + \overline{w}^z \ddy{\overline{u}^z}{z}
                            - f \overline{v}^z & = \overline{F}^z - \overline{D}^z, \\
    \ddy{\overline{v}^z}{t} + \overline{v}^z \ddy{\overline{v}^z}{y}
                            + \overline{w}^z \ddy{\overline{v}^z}{z}
                            + f \overline{u}^z & = - \frac{1}{\rho_0} \ddy{\overline{p}^z}{y}, \\
    0 & = -\frac{1}{\rho_0} \ddy{\overline{p}^z}{z} - \frac{g\overline{\rho}^\rho}{\rho_0}. \label{eqn:mean_hydrostatic}
  \end{align}
\end{subequations}\end{linenomath*}
Here contributions from Reynolds stresses are neglected, it is assumed that all
significant forcing $\overline{F}^z$ and dissipation $\overline{D}^z$ occurs in
the zonal mean momentum equation, and $\overline{\rho}^\rho$ is used in place of
$\overline{\rho}^z$ in the hydrostatic relation (consistent with the discussion
in appendix B of \citealt{McDougallMcIntosh01}). A budget for the mean energy
may be obtained by multiplying by the mean velocity, integrating over the
domain, using incompressibility and the mean density equation, and assuming that
the normal components of both $\overline{\ub}^z$ and $\ub^*$ vanish on all
boundaries. This leads to
\begin{linenomath*}\begin{equation}\label{eqn:mean_energy}
  \frac{\mathrm{d}}{\mathrm{d} t} \iint \left[ \frac{1}{2} \rho_0 \overline{u}^z \overline{u}^z + \frac{1}{2} \rho_0 \overline{v}^z \overline{v}^z + \overline{\rho}^\rho g z \right]\, \mathrm{d}y\, \mathrm{d}z
    = \iint \rho_0 \overline{u}^z \left( \overline{F}^z - \overline{D}^z \right)\, \mathrm{d}y\, \mathrm{d}z + \iint w^* g \overline{\rho}^\rho\, \mathrm{d}y\, \mathrm{d}z.
\end{equation}\end{linenomath*}
The last term is a conversion term which, via substituting $w^*$ from equation
\eqref{eq:u_eddy} and performing an integration by parts, results in
\begin{linenomath*}\begin{equation}\label{eqn:mean_energy_gm}
  \frac{\mathrm{d}}{\mathrm{d} t} \iint \left[ \frac{1}{2} \rho_0 \overline{u}^z \overline{u}^z + \frac{1}{2} \rho_0 \overline{v}^z \overline{v}^z + \overline{\rho}^\rho g z \right]\, \mathrm{d}y\, \mathrm{d}z
    = \iint \rho_0 \overline{u}^z \left( \overline{F}^z - \overline{D}^z \right)\, \mathrm{d}y\, \mathrm{d}z - \iint \rho_0 \kappa \frac{M^4}{N^2}\, \mathrm{d}y\, \mathrm{d}z,
\end{equation}\end{linenomath*}
with horizontal and vertical buoyancy frequencies $M$ and $N$ respectively, where
\begin{linenomath*}\begin{equation}
  M^2 = \frac{g}{\rho_0} \left| \frac{\partial \overline{\rho}^\rho}{\partial y} \right|, \qquad N^2 = -\frac{g}{\rho_0}\frac{\partial \overline{\rho}^\rho}{\partial z}.
\end{equation}\end{linenomath*}
The final term in equation \eqref{eqn:mean_energy_gm} is the conversion of eddy
energy to mean energy. It follows that the eddy energy equation takes the form
(see \ref{app:eddy_energy} for a more complete derivation)
\begin{linenomath*}\begin{equation}\label{eqn:eddy-energy-full}
  \frac{\mathrm{d}}{\mathrm{d} t} \iint \rho_0 E\, \mathrm{d}y\, \mathrm{d}z
    = \iint \rho_0 \kappa \frac{M^4}{N^2}\, \mathrm{d}y\, \mathrm{d}z - \Lambda,
\end{equation}\end{linenomath*}
where $\rho_0 E$ is the eddy energy density, and $\Lambda$ is the dissipation of
eddy energy, for example via topographic form stress. A simple model for this
term is
\begin{linenomath*}\begin{equation}
  \Lambda = -\lambda \iint \rho_0 E\, \mathrm{d}y\, \mathrm{d}z,
\end{equation}\end{linenomath*}
where $\lambda$ is a dissipation time scale. The eddy energy budget
\eqref{eqn:eddy-energy-full} then becomes
\begin{linenomath*}\begin{equation}\label{eqn:eddy_energy}
  \frac{\mathrm{d}}{\mathrm{d} t} \iint E\, \mathrm{d}y\, \mathrm{d}z
    = \iint \kappa \frac{M^4}{N^2}\, \mathrm{d}y\, \mathrm{d}z - \lambda \iint E\, \mathrm{d}y\, \mathrm{d}z.
\end{equation}\end{linenomath*}
The first right-hand-side term in equation \eqref{eqn:eddy-energy-full}, which
is a stratification weighted integral of the GM coefficient, is a consequence of
the GM scheme but is independent on the precise variant of the GM coefficient
used.


\subsection{\cite{Marshall-et-al12} geometric framework and consequences}\label{sec:mmb}

In \cite{Marshall-et-al12} a geometric framework for the eddy fluxes is
proposed. In particular a horizontally down-gradient closure for the horizontal
eddy buoyancy fluxes yields
\begin{linenomath*}\begin{equation}\label{eqn:mmb}
  \kappa = \alpha E \frac{N}{M^2},
\end{equation}\end{linenomath*}
where $\alpha$ is a non-dimensional geometric eddy efficiency parameter that is
bounded in magnitude by one. Provided $\alpha N / M^2$ is bounded away from zero
and infinity, this implies that the magnitude of the GM coefficient should scale
with the eddy energy $E$. This is the case if the mean density has a non-trivial
gradient in both the horizontal and vertical directions, and if the geometric
parameter $\alpha$ is bounded away from zero. Note that the dependence on the
eddy energy is linear, as opposed to a square root dependence that is suggested
by a mixing-length based argument \citep[e.g.,][]{Cessi08, EdenGreatbatch08}.
With this form, once information about the eddy energy is known, for example
from the solution of a parameterised eddy energy budget, then the only remaining
freedom is in the specification of the non-dimensional geometric parameter
$\alpha$.

The physical implications of this closure are described in
\cite{Marshall-et-al16}. Here we highlight the relevant properties in terms of
the expected scaling of the eddy energy on $\alpha$ and the dissipation, and
further the implications for the scaling of the emergent zonal transport, eddy
energy and GM coefficient.

With this GM variant the eddy energy budget \eqref{eqn:eddy_energy} becomes
\begin{linenomath*}\begin{equation}
  \frac{\mathrm{d}}{\mathrm{d} t} \iint E\, \mathrm{d}y\, \mathrm{d}z
    = \iint \left( \alpha \left| s \right| N - \lambda \right) E\, \mathrm{d}y\, \mathrm{d}z,
\end{equation}\end{linenomath*}
where $s = - M^2 / N^2$. In particular, in steady state, the balance
\begin{linenomath*}\begin{equation}
  \iint \left( \alpha \left| s \right| N - \lambda \right) E\, \mathrm{d}y\, \mathrm{d}z = 0
\end{equation}\end{linenomath*}
holds. Note that, from thermal wind shear,
\begin{linenomath*}\begin{equation}\label{eq:thermal-wind}
  \left| \ddy{\overline{u}^z}{z} \right| = \frac{1}{\left| f \right|} \left| s \right| N^2,
\end{equation}\end{linenomath*}
and hence
\begin{linenomath*}\begin{equation}\label{eqn:integral-bal}
  \iint \left( \alpha \frac{\left| f \right|}{N} \left| \ddy{\overline{u}^z}{z} \right| - \lambda \right) E\, \mathrm{d}y\, \mathrm{d}z = 0.
\end{equation}\end{linenomath*}
This is an expression of an eddy energy weighted balance between the eddy energy
generation rate due to the eddies, given by the first integrand term,
and the eddy energy dissipation rate, given by the second. The integral balance
can be achieved if the vertical shear scales with the dissipation rate
$\lambda$, and scales inversely with the geometric parameter $\alpha$. Note
that, following the argument of \cite{Straub93}, the zonal transport scales with
the vertical shear appearing as a factor in the first integrand term. Hence this
suggests that the transport may scale with the dissipation rate $\lambda$, and
scale inversely with the geometric parameter $\alpha$.

For an appropriately smooth eddy energy the following scaling (see
\ref{app:details} for details)
\begin{linenomath*}\begin{equation}\label{eq:slope-constraint}
   \sqrt{\iint \left( \frac{|f|}{N} \left| \ddy{\overline{u}^z}{z} \right| \right)^2 \mathrm{d}y\, \mathrm{d}z} \sim \frac{\lambda}{\alpha}
\end{equation}\end{linenomath*}
is further suggested, again indicating increased transport with increasing
$\lambda$, and decreased transport with increased $\alpha$, but not on the
external forcing $\overline{F}$.

These scalings may be interpreted as follows. Increasing $\lambda$ means the
emergent eddy generation rate needs to increase to maintain the integral balance
\eqref{eqn:integral-bal}, which is achieved via changes in the emergent
stratification profile. This results in steeper isopycnals and thus we expect a
larger transport. An analogous explanation for creasing $\alpha$ suggests an
increase in the transport. 

Analogous scalings of the emergent eddy energy and GM coefficient may be
derived. Consider the mean energy equation along with \eqref{eqn:mmb}. At steady state
and assuming the dissipation of the mean is small, this results in
\begin{linenomath*}\begin{equation}
   \iint \left( \alpha E \frac{|f|}{N} \left| \ddy{\overline{u}^z}{z} \right| 
    - \overline{F}\overline{u}^z \right) \mathrm{d}y\, \mathrm{d}z = 0.
\end{equation}\end{linenomath*}
For fixed $\lambda$, the intrinsic variables $\overline{u}^z$ and $N$ do not
depend on the forcing parameter, so this results in the scaling that $E\sim
\overline{F}$. As a consequence, since $\kappa = \alpha E (N/M^2)$, but $N/M^2$
is large invariant to changes in forcing, this results in $\kappa\sim
\overline{F}$. On the other hand, the functional dependence of the emergent eddy
energy and GM coefficient on varying $\lambda$ and $\alpha$ is not so straight
forward, since it is the vertical stratification weighted transport that is set
by these two parameters. It may be expected that the emergent $\kappa = \alpha E
(N/M^2)$ decreases with increasing $\lambda$ and decreasing $\alpha$, but
primarily through changes in the stratification; the dependence of the eddy
energy level is not obvious.

The suggested dependencies and scalings for the emergent properties are then:
(i) transport to be independent of the magnitude of forcing, increasing with
increased dissipation and decreasing with increased $\alpha$; (ii) GM
coefficient $\kappa$ to scale linearly with the magnitude of wind forcing,
decreasing with increased dissipation and increases with increased $\alpha$;
(iii) eddy energy level to increase linearly with the magnitude of wind forcing.
These are confirmed later via diagnosing the emergent properties from the
simulation data.


\section{Numerical implementation}\label{sec:model}

The \cite{Marshall-et-al12} variant for the GM coefficient given by equation
\eqref{eqn:mmb}, together with the parameterised eddy energy budget in equation
\eqref{eqn:eddy_energy}, is implemented in a simplified two-dimensional model,
similar to that employed in \cite{Gent-et-al95}. The channel model is described
in \S\ref{sec:channel}, and the other parameterisation variants to be tested
against the \cite{Marshall-et-al12} variant are detailed in \S\ref{sec:params}.


\subsection{Channel model}\label{sec:channel}

Employing a linear equation of state, with $\rho / \rho_0 = \beta_s S -
\beta_\theta \theta$, where $\theta$ is temperature, $S$ is salinity, and
$\beta_{s,\theta}$ are expansion coefficients, the prognostic equation for the
mean density is\footnote{Hereafter, as per \citet{McDougallMcIntosh01}, the
density is interpreted as a zonal average on density surfaces, and other
quantities are interpreted as zonal averages at fixed height.}
\begin{linenomath*}\begin{equation}\label{eq:prognostic}
  \ddy{\rhobar}{t} + \ddy{}{y}\left(v^\dagger \rhobar\right) 
    + \ddy{}{z}\left(w^\dagger \rhobar\right) = 0,
\end{equation}\end{linenomath*}
where the mean is taken to be ae zonal average for simplicity. The mean density
is advected by the residual velocity $\ub^\dagger = (0, v^\dagger,
w^\dagger)^T$, which is the sum of the Eulerian circulation $\overline{\ub}$ and
the eddy induced transport velocity $\ub^*$. The domain is chosen to be $\left(
y, z \right) \in (0, L_y) \times (0, L_z)$, and the equation is solved with
no-normal-flow boundary conditions $\ub^\dagger\cdot\nb = 0$ on boundaries.
Dissipation (such as Redi diffusion; \citealt{Redi82}) may be included for
numerical stability, but tests have shown this is not required for the present
idealised model. The model is integrated in time until it reaches a steady
state, with the convergence criterion to be defined.

As a simple model for a forced-dissipative configuration, the Eulerian
circulation appearing in the prognostic equation \eqref{eq:prognostic} is taken
to satisfy the $f$-plane steady state equation
\begin{linenomath*}\begin{equation}\label{eq:eulerian}
  -f\overline{v} = \frac{\tau_s - \tau_b}{\rho_0},\qquad
  \ddy{\overline{v}}{y} + \ddy{\overline{w}}{z} = 0,
\end{equation}\end{linenomath*}
and the thermal wind equation \eqref{eq:thermal-wind}, where $\tau_s$ is
the surface wind stress, and $\tau_b$ is a representation of the bottom form
stress \citep[see][]{Marshall97}. The surface wind stress is taken to be
\begin{linenomath*}\begin{equation}
  \tau_s = \frac{\tau_0}{2\Delta z}\left(1 - \cos\frac{2\pi y}{L_y}\right)
\end{equation}\end{linenomath*}
with peak wind stress $\tau_0$. The bottom stress $\tau_b$ is chosen to exactly
cancel the local surface wind, $\tau_b = -\tau_s$. With this choice, the
equation for $\overline{u}$ becomes purely diagnostic. The Eulerian circulation
is thus specified by the forcing $\tau_s$, while the state $\overline{\rho}$
determines the eddy induced transport velocity $(0, v^*, w^*)^T$ through
equation \eqref{eq:u_eddy}, which is then used to form the residual velocity to
time step the prognostic equation \eqref{eq:prognostic}.

The prognostic equations are discretised in space using a uniform resolution
Arakawa C-grid \citep{ArakawaLamb77} with $y$- and $z$-direction grid spacings
of $\Delta y$ and $\Delta z$ respectively. The density $\rhobar$ is defined at
the cell centres, fluxes are defined on the cell interfaces, and derivatives of
$\rhobar$ on cell corners, with appropriate interpolation of the fields where
required. The boundary conditions are implemented by setting boundary fluxes to
zero. The forcing and dissipation are taken to be applied over the top and
bottom cells. A fourth order Runge--Kutta method is employed to time step the
prognostic equation \eqref{eq:prognostic} and eddy energy equation
\eqref{eqn:eddy_energy}, with a variable $\Delta t$ chosen at the end of each
time step so as to target a desired Courant number $C$ \citep{Courant-et-al28},
\begin{linenomath*}\begin{equation}
  \Delta t = C\left(\frac{v^\dagger}{\Delta y} + \frac{w^\dagger}{\Delta z}\right).
\end{equation}\end{linenomath*}
For numerical stability in integrating the parmaeterised eddy energy equation,
the variable time step is restricted so that $\Delta t \leq 12$ hours. The
calculations are initialised with an exponential density profile
\begin{linenomath*}\begin{equation}\label{eq:rho-spinup-init}
  \overline{\rho}(t = 0) = a + b\mathrm{e}^{z/c}.
\end{equation}\end{linenomath*}

Slope clipping \citep{Cox87} is used to avoid unbounded velocities associated
with weak stratification. The value of the slope $s$ appearing in the
parameterised eddy transport velocity \eqref{eq:u_eddy} is replaced with the
slope clipped value of $\tilde{s}$ 
\begin{linenomath*}\begin{equation} 
  \tilde{s} = \min\left( s_\mathrm{max}, s \right), 
\end{equation}\end{linenomath*}
with a chosen value of the maximum slope $s_\mathrm{max}$ . For noise reduction,
Gaussian smoothing is applied to the slope field $s$. Additionally, in equations
in which a division by the vertical stratification $N^2$ appears (e.g. the first
right-hand-side term of equation \eqref{eqn:eddy_energy}) this is replaced with
\begin{linenomath*}\begin{equation}
  \tilde{N}^2 = \max \left( N_\mathrm{min}^2, N^2 \right)
\end{equation}\end{linenomath*}
with a chosen value of the minimum vertical stratification $N_\mathrm{min}$.

During time stepping a basic convection scheme is applied, with each vertical
water column sorted by density within each Runge--Kutta stage. This facilitates
the development of out-cropping at the surface, which would otherwise be
constrained by the initially constant surface density and the no-normal-flow
boundary condition. The convection scheme is disabled when
\begin{linenomath*}\begin{equation}\label{eq:measure}
  \mathcal{E} = \frac{\int |\rhobar_2 - \rhobar_1|^2\, \mathrm{d}y\,
  \mathrm{d}z}{\int |\rhobar_1|^2\, \mathrm{d}y\, \mathrm{d}z} < \xi_1,
\end{equation}\end{linenomath*}
where $\rhobar_{1,2}$ are outputs that are separated in time by some threshold
(taken to be 50 days in dimensional time), and $\xi_1$ is a user-defined
tolerance. A solution is deemed to have converged to a steady state when
$\mathcal{E} < \xi_2$, for some convergence threshold $\xi_2 < \xi_1$. For each
of the two cases (spatially constant in \S\ref{sec:const-spinup} and
stratification dependent in \S\ref{sec:vert-spinup}) an initial steady state
control run with a wind forcing of $\tau_0 = 0.2\ \mathrm{N}\ \mathrm{m}^{-2}$
was computed, and used as the initial condition for further calculations. These
calculations were each integrated for a maximum of a further $500$ years if
$\tau_0 > 0.1\ \mathrm{N}\ \mathrm{m}^{-2}$, and for a maximum of a further
$1,500$ years if $\tau_0 \leq 0.1\ \mathrm{N}\ \mathrm{m}^{-2}$. If a steady
state was not reached in this time the calculation was excluded from further
analysis; this affects only the stratification dependent case.

Model parameter values are provided in Table~\ref{tbn:model_spinup}.

\begin{table}[tbhp]
  \begin{center}
  {\small
    \begin{tabular}{|c|c|c|l|}
      \hline
      Parameter & Value & Units & Description\\
      \hline
      ($L_y, L_z$) & (2000, 3) & $\mathrm{km}$ 
        & domain size \\
      ($\Delta y, \Delta z)$ & (10, 0.1) & $\mathrm{km}$ 
        & grid spacing\\
      $C$ & 0.1 & --- 
        & CFL number\\
      $s_\mathrm{max}$ & $1\times 10^{-2}$ & --- & 
        slope clipping value $s$ in generating the eddy induced transport velocity\\
      $N_\mathrm{min}^2$ & $5\times 10^{-6}$ & $\mathrm{s}^{-2}$ & 
        minimum value of $N^2$ in the integrands\\
      $\xi_1$ & $10^{-13}$ & --- 
        & tolerance for switching off convective sorting scheme\\
      $\xi_2$ & $10^{-15}$ & --- 
        & tolerance for solution convergence\\
      $f_0$ & $-10^{-4}$ & $\mathrm{rad}$~$\mathrm{s}^{-1}$ 
        & Coriolis parameter \\
      $\rho_0$ & 1000 & $\mathrm{kg}\ \mathrm{m}^{-3}$ 
        & reference density\\
      $g$ & 9.8 & $\mathrm{m}\ \mathrm{s}^{-2}$ 
        & gravitational acceleration\\
      $a$ & 1000 & $\mathrm{kg}\ \mathrm{m}^{-3}$ 
        & base density for $\overline{\rho}(t = 0)$ given in \eqref{eq:rho-spinup-init}\\
      $b$ & $0.6$ & $\mathrm{kg}\ \mathrm{m}^{-3}$ 
        & factor for $\overline{\rho}(t = 0)$ given in \eqref{eq:rho-spinup-init}\\
      $c$ & $750$ & $\mathrm{m}$ 
        & $e$-folding depth for $\overline{\rho}(t = 0)$ given in \eqref{eq:rho-spinup-init}\\
      \hline
    \end{tabular}
  }
  \end{center}
  \caption{Parameter values used in the numerical model.}
  \label{tbn:model_spinup}
\end{table}


\subsection{Alternative GM eddy transfer coefficients}\label{sec:params}

For comparison, a number of alternative variants based on existing
parameterisation schemes are also implemented in the idealised numerical model.
A scheme that employs a mixing length assumption and has dependence on the eddy
energy is given by
\begin{linenomath*}\begin{equation}
  \kappa = \alpha_{\tiny \mbox{ML}} \sqrt{E} L,
\end{equation}\end{linenomath*}
where $\alpha_{\tiny \mbox{ML}}$ is some non-dimensional parameter (without a
formal bound) and $L$ is a mixing length scale to be specified. This scheme has
a weaker dependence on the eddy energy. An approach of this form is described in
\cite{EdenGreatbatch08}, where the eddy energy is replaced with the eddy kinetic
energy, and the length scale is taken to be the minimum of the Rhines scale and
the Rossby deformation radius (their equation 25). Setting the mixing length
equal to the Rhines scale increases the eddy kinetic energy exponent to $3/4$,
and hence this is closer to the linear energy scaling in equation
\eqref{eqn:mmb}. A similar mixing length approach is taken in \cite{Cessi08}
where a statistically steady version of \eqref{eqn:eddy_energy} is utilised to
derive a form of the GM coefficient that has explicit dependence on the bottom
drag. \cite{Cessi08} uses the eddy kinetic energy in place of the eddy energy,
and choose $L$ to be the Rossby deformation radius.

Note that the derivation of \cite{EdenGreatbatch08}, in their equation (26),
suggests that the GM coefficient should have a linear dependence on the eddy
kinetic energy. However in their work the chosen length scale implicitly sets
the magnitude of the eddy kinetic energy. Here, instead, the eddy energy is
parameterised directly. In \cite{Jansen-et-al15} a mixing length which scales
with the square root of the eddy kinetic energy is discussed, yielding a form
equivalent to the scaling of \eqref{eqn:mmb}, with the eddy kinetic energy again
used in place of the eddy energy.

Based on instability arguments, \cite{Visbeck-et-al97} proposed
\begin{linenomath*}\begin{equation}\label{eq:vmhs-raw}
  \kappa = \alpha_{\tiny \mbox{VMHS}} \frac{L^2}{T} = \alpha_{\tiny \mbox{VMHS}} L^2 \frac{M^2}{N},
\end{equation}\end{linenomath*}
where $\alpha_{\tiny \mbox{VMHS}}$ is some non-dimensional parameter (again
without a formal bound). This variant has no explicit dependence on the eddy
energy, and instead depends only on the mean stratification. In \S3$d$ of
\cite{Visbeck-et-al97} the length scale $L$ is related to the grid scale, Rossby
deformation radius, and the width of the baroclinic zone.

Diagnosing diffusivities from a high resolution numerical ocean model
constrained using observation data and via an adjoint based optimisation,
\cite{Ferreira-et-al05} instead suggested that
\begin{linenomath*}\begin{equation}\label{eqn:structure}
  \kappa = \kappa_0 \mathcal{S}, \qquad \mathcal{S} = \frac{N^2}{N^2_{\textnormal{ref}}},
\end{equation}\end{linenomath*}
where $\kappa_0$ is some reference GM coefficient value, and $\mathcal{S}$
imparts a spatial structure to the GM coefficient that is dependent on the
vertical stratification. This results in a GM coefficient that is large towards
the ocean surface whilst being small in the deep ocean where the stratification
is weak. The reference value $\kappa_0$ is normally taken to be constant
\citep[e.g.,][]{Ferreira-et-al05, DanabasogluMarshall07, GentDanabasoglu11}.


\subsection{Summary}\label{sec:param-sum}

In summary, the four variants for the GM eddy transfer coefficient considered in
this article are:
\begin{itemize}
  \item a constant GM coefficient, denoted CONST;
  \item the \cite{Marshall-et-al12} derived variant, denoted GEOM;
  \item a mixing length variant similar to the approach of
  \cite{EdenGreatbatch08} and \cite{Cessi08}, denoted ML;
  \item a scheme similar to that described in \cite{Visbeck-et-al97}, denoted
  VMHS${}^*$.
\end{itemize}
Each of these four variants are considered subject to two approximations, with
implementation details given in the appropriate sections. This first is where
the GM coefficient is assumed to be spatially constant. The second is one where
the GM coefficient has an imposed spatial structure set by $\mathcal{S}$ from
equation \eqref{eqn:structure}, to be in line with more modern numerical models
\citep[e.g.,][]{DanabasogluMarshall07, GentDanabasoglu11}. Where relevant all
length scales are set equal to the Rossby deformation radius. All the
implemented variants are coupled to the parameterised eddy energy equation
\eqref{eqn:eddy_energy}, although this plays a prognostic role only for the GEOM
and ML variants.

The use of a prescribed spatial structure for the GM coefficient contradicts
somewhat with the original intention of the scheme described in
\cite{Visbeck-et-al97}. Thus a variant, denoted VMHS, is additionally
considered, which uses the full local dependence as specified in equation
\eqref{eq:vmhs-raw}. Note, however, the length scale is still set equal to the
Rossby deformation radius, which differs from the length scale used in
\cite{Visbeck-et-al97}. 

The four parameterisation variants differ in their functional dependence on the
eddy energy and the mean stratification, as summarised in
Table~\ref{tbn:param-variant}.

\begin{table}[tbhp]
  \begin{center}
  {\small
    \begin{tabular}{|c|l|c|c|c|}
      \hline
      Variant & Functional form & A & B & C \\
      \hline
      CONST
        & $\kappa = \kappa_0 ~~~~~~~~~~~~~~~ = \kappa_0 E^0 M^0 N^0$
        & $0$ & $0$ & $0$ \\
      \hline
      GEOM
        & $\kappa = \alpha E T ~~~~~~~~~~ = \alpha E^1 M^{-2} N^1$
        & $1$ & $-2$ & $1$ \\
      \hline
      ML
        & $\kappa = \alpha_{\tiny \mbox{ML}} \sqrt{E} L_D ~ = \alpha_{\tiny \mbox{ML}} (H / f_0) E^{1/2} M^0 N^1$
        & $1/2$ & $0$ & $1$ \\
      \hline
      VMHS${}^*$ and VMHS
        & $\kappa = \alpha_{\tiny \mbox{VMHS}} L_D^2 / T = \alpha_{\tiny \mbox{VMHS}} (H^2 / f_0^2) E^0 M^2 N^1$
        & $0$ & $2$ & $1$ \\
      \hline
    \end{tabular}
  }
  \end{center}
  \caption{Functional dependence of the four considered variants on the 
  eddy energy $E$, the horizontal stratification $M$, and the vertical
  stratification $N$, expressed in the form $\kappa \propto E^A M^B N^C$. Where
  relevant the mixing length parameter has been set equal to the Rossby
  deformation radius $L_D = N H / f_0$.}
  \label{tbn:param-variant}
\end{table}


\section{Spatially constant Gent--McWilliams coefficient}\label{sec:const-spinup}

\subsection{Implementation details}

In this section the case of spatially constant GM coefficient is considered,
employing the CONST, GEOM, ML and VMHS${}^*$ variants described in
\S\ref{sec:param-sum}. The CONST variant is simply employed by taking a constant
value of $\kappa$. To obtain a spatially constant GM coefficient for the GEOM
variant with $\kappa = \alpha E (N/M^2)$, the terms are appropriately
re-arranged, and integrating over the domain leads to
\begin{linenomath*}\begin{equation}
  \kappa = \alpha \frac{\iint E\, \mathrm{d}y\,\mathrm{d}z}{\iint (M^2 / N)\, \mathrm{d}y\, \mathrm{d}z}.
\end{equation}\end{linenomath*}
The domain integrated eddy energy $\iint E\, \mathrm{d}y\,\mathrm{d}z$ is
computed by solving \eqref{eqn:eddy_energy}.

For the ML variant, an analogous approach yields
\begin{linenomath*}\begin{equation*}
  \kappa = \alpha_{\tiny \mbox{ML}} \frac{\iint \sqrt{E}\, \mathrm{d}y\,\mathrm{d}z}{\iint (1 / L)\, \mathrm{d}y\,\mathrm{d}z}.
\end{equation*}\end{linenomath*}
However the domain integral of the square root of the eddy energy is not
available. Use of the Cauchy--Schwarz inequality (e.g., \ref{app:details} with
$p=q=2$) leads to
\begin{linenomath*}\begin{equation*}
  \iint \sqrt{E}\, \mathrm{d}y\,\mathrm{d}z \leq \sqrt{L_y L_z} \sqrt{\iint E\, \mathrm{d}y\,\mathrm{d}z},
\end{equation*}\end{linenomath*}
and so the ML variant is implemented as
\begin{linenomath*}\begin{equation}\label{eq:eg}
  \kappa = \alpha_1 \frac{\sqrt{L_y L_z} \sqrt{\iint E\, \mathrm{d}y\,\mathrm{d}z}}{\iint (1 / L)\, \mathrm{d}y\,\mathrm{d}z}.
\end{equation}\end{linenomath*}
Here a prescribed value for the new parameter $\alpha_1$ is chosen.

For the VMHS${}^*$ variant the form
\begin{linenomath*}\begin{equation}\label{eq:vmhs-star}
  \kappa = \alpha_2 \frac{\iint L^2 (M^2 / N)\, \mathrm{d}y\,\mathrm{d}z}{L_y L_z},
\end{equation}\end{linenomath*}
is used. The VMHS variant with fully local dependence on the mean state is
considered in \S\ref{sec:vert-spinup}.

\begin{figure}
  \includegraphics[width=\linewidth]{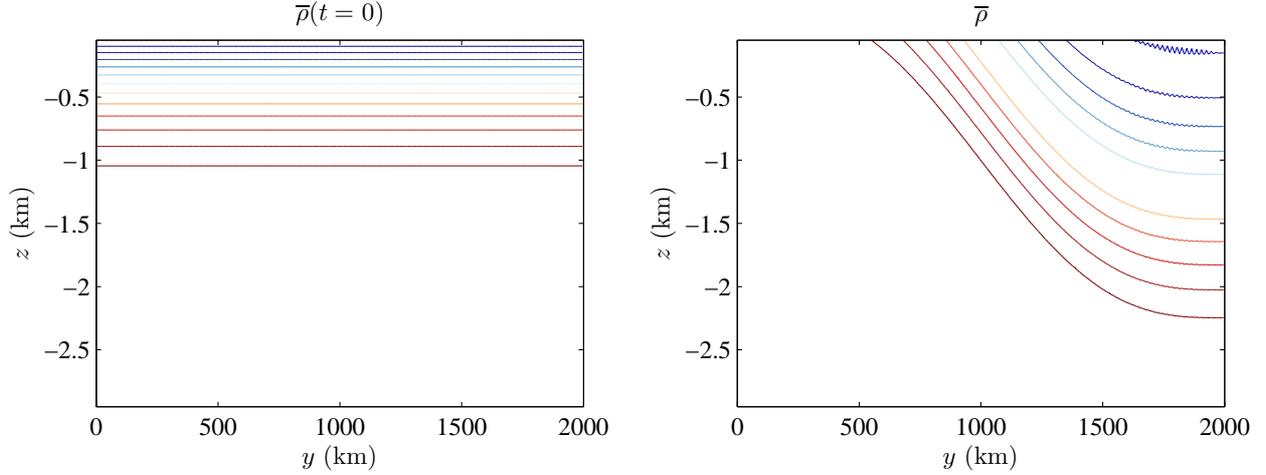}
  \caption{Initial stratification and equilibrium stratification from the spinup
  of the control run with $\alpha = 0.1$, $\tau_0 = 0.2\ \mathrm{N}\
  \mathrm{m}^{-2}$ and $\lambda = 2\times10^{-7}\ \mathrm{s}^{-1}$ for the GEOM
  variant (leading to an emergent $\kappa = 805\ \mathrm{m}^{2}
  \mathrm{s}^{-1}$). The same contour levels are used for both panels.}
  \label{fig:spinup_init_final}
\end{figure}

The initial state is spun up from rest first using the GEOM variant, with
$\tau_0 = 0.2\ \mathrm{N}\ \mathrm{m}^{-2}$, $\lambda = 2\times10^{-7}\
\mathrm{s}^{-1}$ and $\alpha = 0.1$. The associated initial and equilibrium
states are shown in Figure~\ref{fig:spinup_init_final}. The equilibrium state
here has a transport of around 77 $\mathrm{Sv}$ and a domain average
parameterised eddy energy of around $0.01\ \mathrm{m}^2\ \mathrm{s}^{-2}$, the
latter being similar to the level given in the observations of
\cite{MeredithHogg06}. From this control run and taking the mixing length $L$ to
be the Rossby deformation radius $L_d = NH / f_0$ for the ML and VMHS${}^*$
variants, the emergent $\kappa$ and end state $\overline{\rho}$ are used to
calibrate $\kappa$ for CONST, $\alpha_1$ for the ML variant in \eqref{eq:eg} and
$\alpha_2$ for the VMHS${}^*$ variant in \eqref{eq:vmhs-star}, which are used
for subsequent calculations where $\tau_0$ and $\lambda$ are varied. The
resulting emergent GM coefficients, eddy energy, and the mean transport are
computed.

\subsection{Results}\label{sec:const-results}

\begin{figure}
  \includegraphics[width=\linewidth]{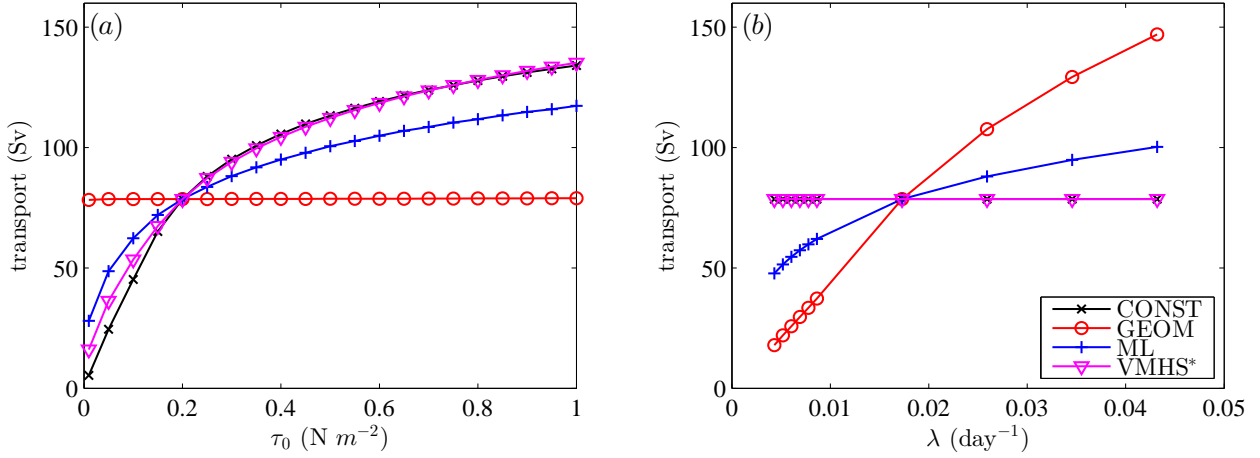}
  \caption{Transport at varying ($a$) peak wind forcing $\tau_0$ and ($b$)
  eddy energy dissipation rate $\lambda$ for the four parameterisation
  variants.}
  \label{fig:transport_varyparam}
\end{figure}

The transport associated with the equilibrium states with varying values for
$\tau_0$ and $\lambda$ are shown in Figure~\ref{fig:transport_varyparam}. It is
clear that CONST and VMHS${}^*$ show significant sensitivity of the mean
transport with respect to the peak wind stress. By contrast, the ML variant
shows reduced sensitivity. Notably, the GEOM variant shows very low sensitivity
to varying wind stress, and thus exhibits the emergent eddy saturation, obtained
in the eddy-permitting numerical experiments of \cite{Munday-et-al13} for
example. For varying eddy energy dissipation rate $\lambda$, CONST and
VMHS${}^*$ are by construction independent of $\lambda$, while the ML and GEOM
variants show increased transport with increased dissipation. These observed
behavious are consistent with the analysis given in \S\ref{sec:mmb}.

\begin{figure}
  \includegraphics[width=\linewidth]{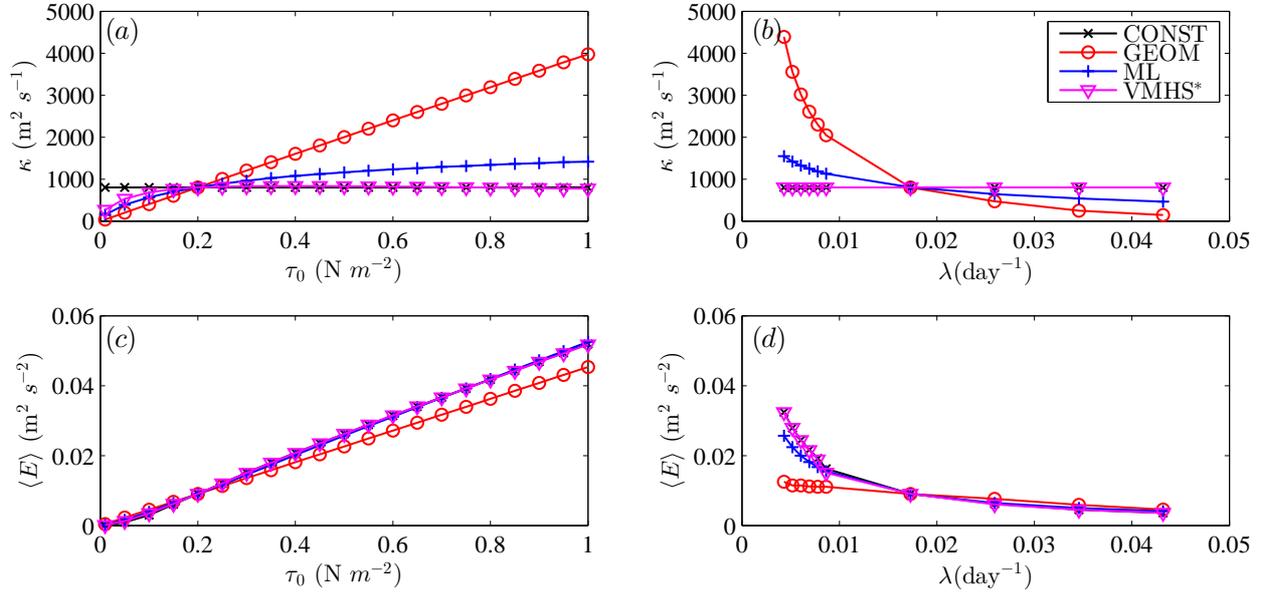}
  \caption{Emergent ($a,b$) $\kappa$ and ($c,d$) domain integrated eddy energy
  $\langle E\rangle$ at varying peak wind forcing $\tau_0$ ($a,c$) and
  eddy energy dissipation rate $\lambda$ ($b,d$), for the four parameterisation
  variants.}
  \label{fig:kappa_E_varyparam}
\end{figure}

Denoting the domain average by
\begin{linenomath*}\begin{equation}
  \langle\cdot\rangle = \frac{1}{L_y L_z}\iint (\cdot)\, \mathrm{d}y\, \mathrm{d}z,
\end{equation}\end{linenomath*}
the emergent $\kappa$ and $\langle E\rangle$ are shown in
Figure~\ref{fig:kappa_E_varyparam}. The ML and VMHS${}^*$ variants show a
sub-linear dependence of the emergent $\kappa$ on the peak wind stress $\tau_0$,
while the GEOM variant exhibits an almost linear dependence. For the ML and GEOM
variants the emergent $\kappa$ decreases with increasing $\lambda$. The
dependencies are consistent with the arguments given in \S\ref{sec:mmb}. It is
found here that increasing the dissipation decreases the emergent eddy energy
level.

\begin{figure}
  \includegraphics[width=\linewidth]{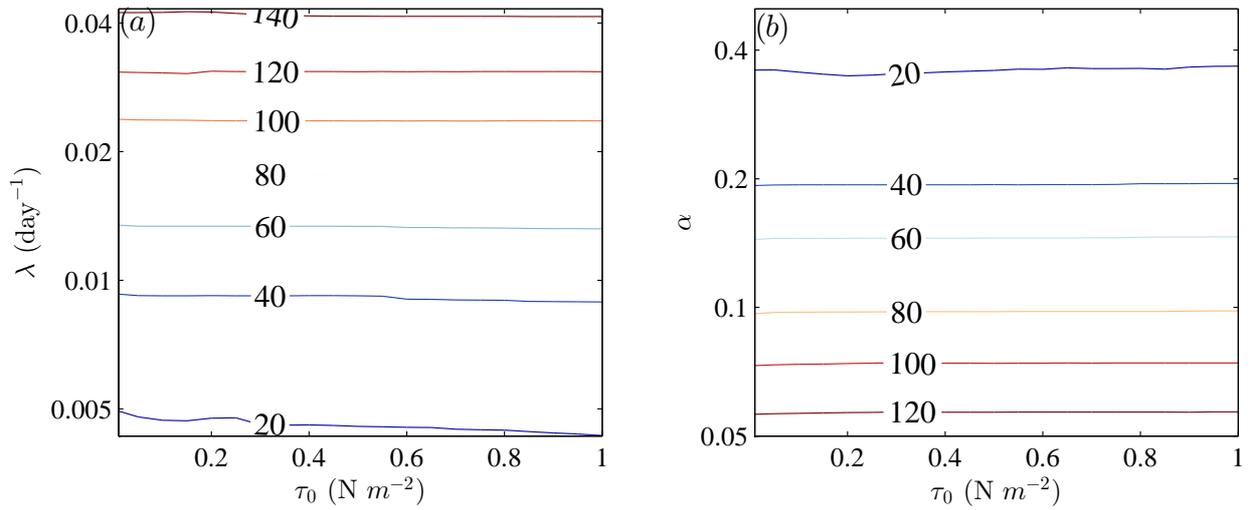}
  \caption{Contour plot of the emergent transport (in Sv) for the GEOM variant
  over ($a$) ($\tau_0, \lambda$) space (with $\alpha = 0.1$), and ($b$)
  ($\tau_0, \alpha$) space (with $\lambda = 2\times 10^{-7}\ \mathrm{s}^{-1}$).}
  \label{fig:transport_contour}
\end{figure}

The emergent eddy saturation property of the GEOM variant is not limited to this
parameter set. Figure~\ref{fig:transport_contour} shows contour plots of the
transport in $(\tau_0, \lambda)$ and $(\tau_0, \alpha)$ parameter space. As
expected, there is very little dependence of the transport on $\tau_0$ and only
at extreme parameter values is a variability seen in the contour plot. This
shows robustness of the insensitivity to strength of peak wind over a large
range of parameters.

\begin{figure}
  \includegraphics[width=\linewidth]{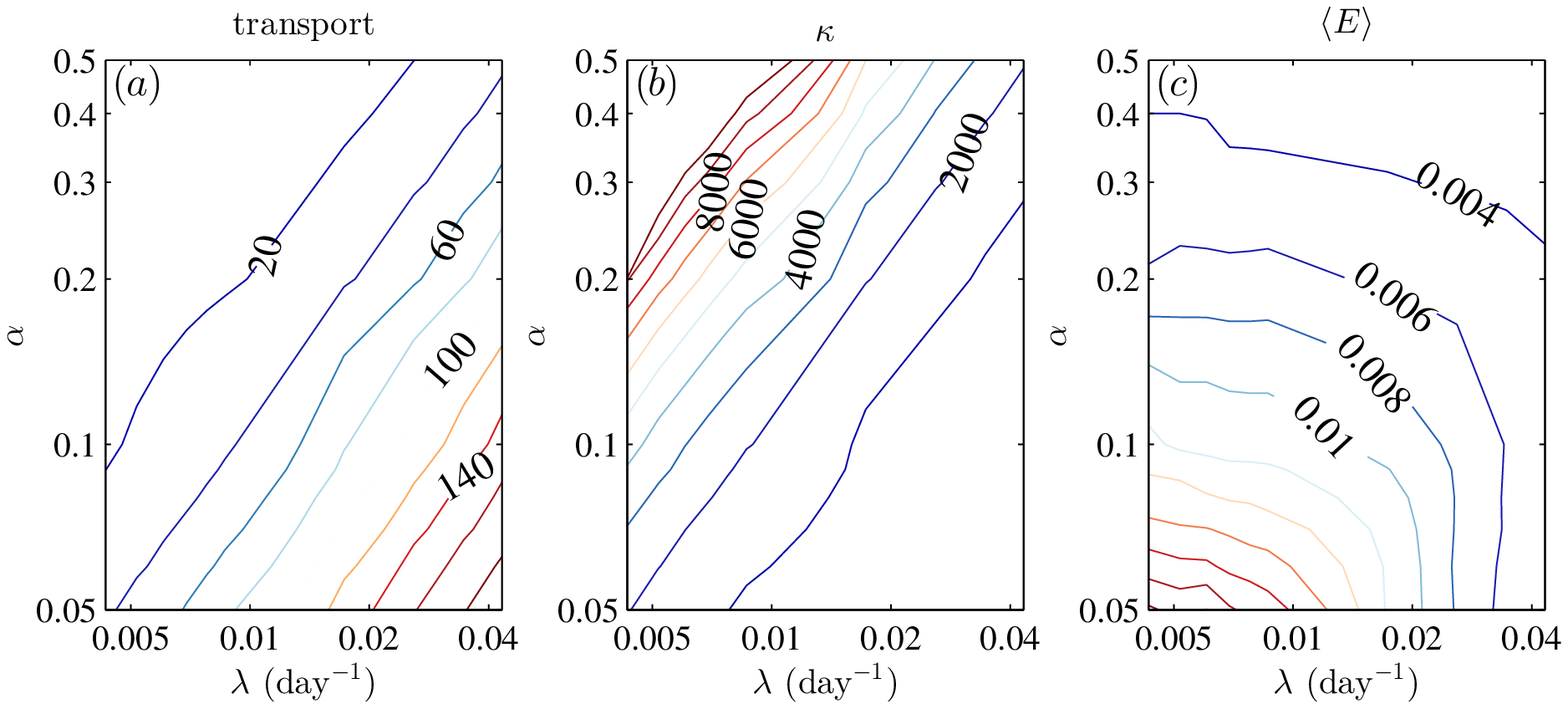}
  \caption{Contour plot of the emergent ($a$) transport (in Sv), ($b$) $\kappa$
  (in $\ \mathrm{m}^2\ \mathrm{s}^{-1}$) and ($c$) $\langle E \rangle$ (in $\
  \mathrm{m}^2\ \mathrm{s}^{-2}$) of the GEOM variant over $\lambda$ and
  $\alpha$ space, at $\tau_0 = 0.2\ \mathrm{N}\ \mathrm{m}^{-2}$. The axes are
  logarithmic with simulation data from 10 values of $\alpha$ ranging from
  $0.05$ to $0.5$ and 10 values of $\lambda$ ranging from $5\times 10^{-8}\
  \mathrm{s}^{-1}$ to $5\times 10^{-7}\ \mathrm{s}^{-1}$.}
  \label{fig:emergent_contour}
\end{figure}

To show how the other emergent properties of the GEOM variant depend on $\alpha$
and $\lambda$, the transport, GM coefficient, and domain averaged eddy energy
over $(\lambda, \alpha)$ parameter space are shown in
Figure~\ref{fig:emergent_contour}. Increasing $\alpha$ increases $\kappa$ and
reduces the mean transport as expected. However, the dependence of $\kappa$ on
$\alpha$ is not linear, due to the indirect effect of $\alpha$ on the eddy
energy. For lower values of $\alpha$, decreasing $\lambda$ leads to an increase
in the eddy energy, an increase in $\kappa$, and a decrease in the mean
transport. The eddy energy has a more complex dependence on $\alpha$, but for
weaker dissipation increasing $\alpha$ leads to a decrease in the eddy energy.
Again, these observations are consistent with the analysis given in
\S\ref{sec:mmb}.

\section{Stratification dependent Gent--McWilliams coefficient}\label{sec:vert-spinup}

\subsection{Implementation details}

In this section a dependence of the GM coefficient on the vertical
stratification is introduced, again with four variants based upon the CONST,
GEOM, ML, and VMHS${}^*$ discussed in \S\ref{sec:params}. The simplest CONST
variant is now replaced with the form proposed in \cite{Ferreira-et-al05}
\begin{linenomath*}\begin{equation}\label{eq:const-space}
  \kappa = \kappa_0 \mathcal{S},\qquad 
  \mathcal{S} = \frac{N^2}{N^2_{\textnormal{ref}}}.
\end{equation}\end{linenomath*}
This imparts a vertical as well as horizontal spatial structure to the GM
coefficient. In \cite{Ferreira-et-al05} $N^2_{\textnormal{ref}}$ is taken to be
the value of $N^2$ at the surface, and the $\mathcal{S}$ is tapered to a value
of $1$ to avoid singularities (for example, during a convective event when
outcropping occurs). Here this was found to lead to difficulties in regions of
weak surface stratification, which were not alleviated by the use of clipping of
$\mathcal{S}$. Instead, here a simpler approach is taken, with
$N^2_{\textnormal{ref}}$ set equal to a constant value, for convenience set
equal to $N^2_{\textnormal{min}}$ (see Table~\ref{tbn:model_spinup}).

The GEOM variant becomes $\kappa = \kappa_0 \mathcal{S} = \alpha E (N/M^2)$,
where again $\alpha$ is a prescribed constant. Re-arranging, integrating over
the domain, and now assuming that $\kappa_0$ is a constant in space leads to
\begin{linenomath*}\begin{equation}\label{eq:kappa_space}
  \kappa = \kappa_0 \mathcal{S} = 
  \left(\alpha \frac{\iint E\, \mathrm{d}y\, \mathrm{d}z}
    {\iint (M^2 / N) \mathcal{S}\, \mathrm{d}y\, \mathrm{d}z}\right)
    \mathcal{S}.
\end{equation}\end{linenomath*}
The domain integrated eddy energy $\iint E\, \mathrm{d}y\, \mathrm{d}z$ is
computed by solving equation \eqref{eqn:eddy_energy} as before.

For the ML variant, an analogous approach yields
\begin{linenomath*}\begin{equation*}
  \kappa = \left( \alpha_{\tiny \mbox{ML}} \frac{\iint \sqrt{E}\, \mathrm{d}y\, \mathrm{d}z}{\iint (\mathcal{S} / L)\, \mathrm{d}y\, \mathrm{d}z} \right) \mathcal{S},
\end{equation*}\end{linenomath*}
and use of the Cauchy--Schwarz inequality leads to
\begin{linenomath*}\begin{equation}\label{eq:eg-space}
  \kappa = \left( \alpha_1 \frac{\sqrt{L_y L_z} \sqrt{\iint E\, \mathrm{d}y\, \mathrm{d}z}}{\iint (\mathcal{S} / L)\, \mathrm{d}y\, \mathrm{d}z} \right) \mathcal{S}.
\end{equation}\end{linenomath*}
Here a prescribed value for the parameter $\alpha_1$ is again chosen.

For the variant based on \cite{Visbeck-et-al97}, with the GM coefficient given
by $\kappa = \kappa_0 \mathcal{S} = \alpha_{\tiny \mbox{VMHS}} (M^2/N) L^2$, two
forms are used. Assuming $\kappa_0$ is a constant in space results in the
VMHS${}^*$ variant
\begin{linenomath*}\begin{equation}\label{eq:vmhs-space}
  \kappa = \left(\alpha_2 \frac{\iint L^2 (M^2 / N)\, \mathrm{d}y\, \mathrm{d}z}
    {\iint \mathcal{S}\, \mathrm{d}y\, \mathrm{d}z}\right)
  \mathcal{S}.
\end{equation}\end{linenomath*}
Alternatively the form \eqref{eq:vmhs-raw} may be used directly, resulting in
the VMHS variant
\begin{linenomath*}\begin{equation}\label{eq:vmhs}
  \kappa = \alpha_3 \frac{M^2}{N} L^2,
\end{equation}\end{linenomath*}
where now $\alpha_3$ is a prescribed constant. This latter form introduces an
additional explicit dependence on the local value of $M$ and the local mixing
length $L$. 

\begin{figure}
  \includegraphics[width=\linewidth]{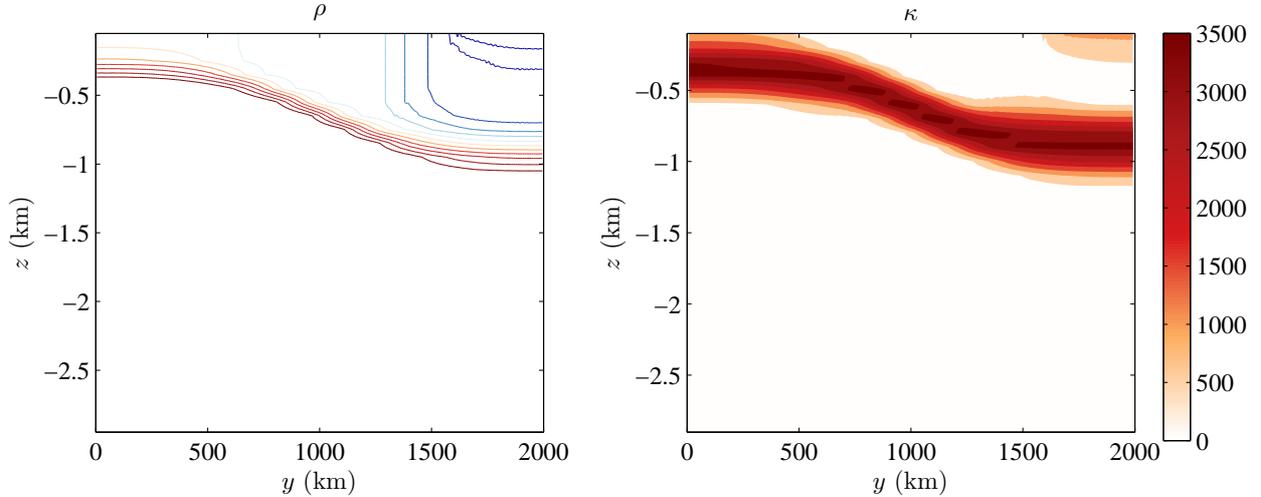}
  \caption{Equilibrium stratification and final GM coefficient distribution from
  control run with $\alpha = 0.1$ and $\lambda = 2\times10^{-7}\
  \mathrm{s}^{-1}$. The contour levels for the stratification profile are the
  same as for Figure~\ref{fig:spinup_init_final}.}
  \label{fig:final_control_spinup}
\end{figure}

As for previous constant GM coefficient case, the initial state is spun up from
rest first using the GEOM variant, with $\tau_0 = 0.2\ \mathrm{N}\
\mathrm{m}^{-2}$, $\lambda = 2\times10^{-7}\ \mathrm{s}^{-1}$ and $\alpha =
0.1$. The initial state is the same one shown in
Figure~\ref{fig:spinup_init_final}, and Figure~\ref{fig:final_control_spinup}
shows the equilibrium stratification profile and the associated spatially
varying GM coefficient. This equilibrium state here has a transport of around 66
$\mathrm{Sv}$ and a domain average parameterised eddy energy of around $0.004\
\mathrm{m}^2\ \mathrm{s}^{-2}$. This energy is somewhat lower than the previous
constant GM coefficient case, possibly due to the anticipated restriction of the
parameterised eddy energy to strongly sloping isopycnals, which are confined to
the upper part of the domain.

From this control run and taking the mixing length $L$ to be the Rossby
deformation radius $L_d = NH / f_0$ as before for the ML and VMHS${}^*$
variants, the emergent $\kappa$ and end state $\overline{\rho}$ are used to
calibrate $\kappa_0$ for CONST in \eqref{eq:const-space}, $\alpha_1$ for ML in
\eqref{eq:eg-space} and $\alpha_2$ for the VMHS${}^*$ in \eqref{eq:vmhs-space},
which are used for subsequent calculations where $\tau_0$ and $\lambda$ are
varied. For the direct VMHS variant, the functional dependence of $\kappa$ on
$M$ and $N$ differs from the functional dependence specified by $\mathcal{S}$.
Here an initial value of $\alpha_3$ in \eqref{eq:vmhs} is chosen manually so
that a similar transport is obtained. Other simulation details are kept the same
as in Table~\ref{tbn:model_spinup} except that the convergence tolerance $\xi_2$
is now set to $5\times10^{-14}$, as there is more variability given that
$\kappa$ is allowed to vary over space.

\subsection{Results}\label{sec:vert-results}

\begin{figure}
  \includegraphics[width=\linewidth]{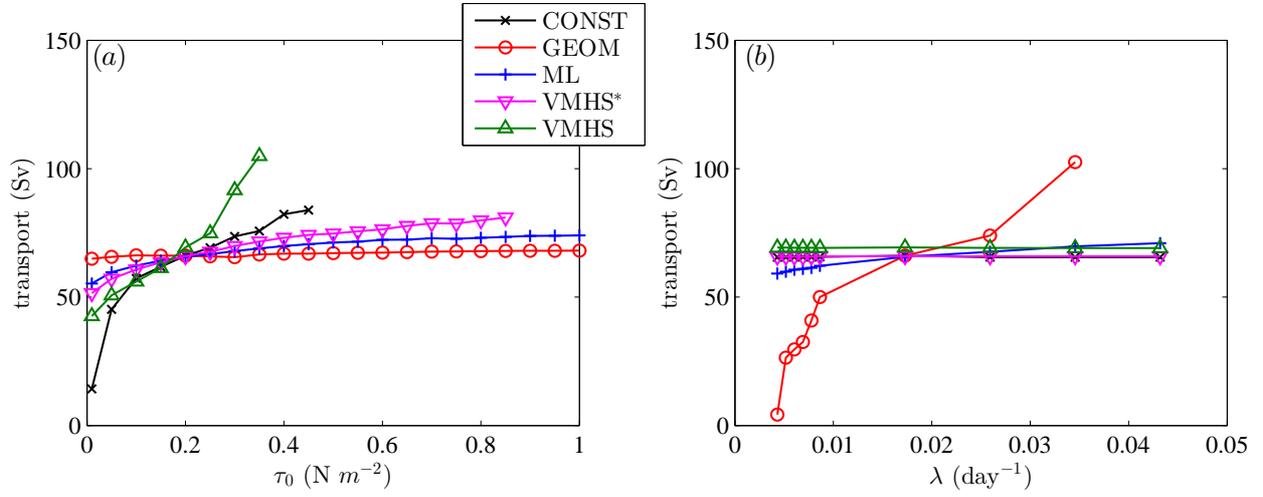}
  \caption{Transport at varying ($a$) peak wind forcing $\tau_0$ and ($b$)
  eddy energy dissipation rate $\lambda$ for the five parameterisation
  variants, showing only converged solutions.}
  \label{fig:transport_varyparam_space}
\end{figure}

The resulting transport with varying $\tau_0$ and $\lambda$ for the five
parameterisation variants is shown in
Figure~\ref{fig:transport_varyparam_space}. The GEOM variant once again exhibits
emergent eddy saturation behaviour. The VMHS${}^*$ variant and especially the ML
variant also show a reduction of the sensitivity of the mean transport with
respect to the peak wind stress compared to their respective case with spatially
constant GM coefficient. The GEOM variant once again exhibits a strong
dependence on the eddy energy dissipation rate $\lambda$. The ML variant with
stratification dependence now exhibits a much weaker dependence of the transport
on $\lambda$ than was observed with a constant GM coefficient case.

\begin{figure}
  \includegraphics[width=\linewidth]{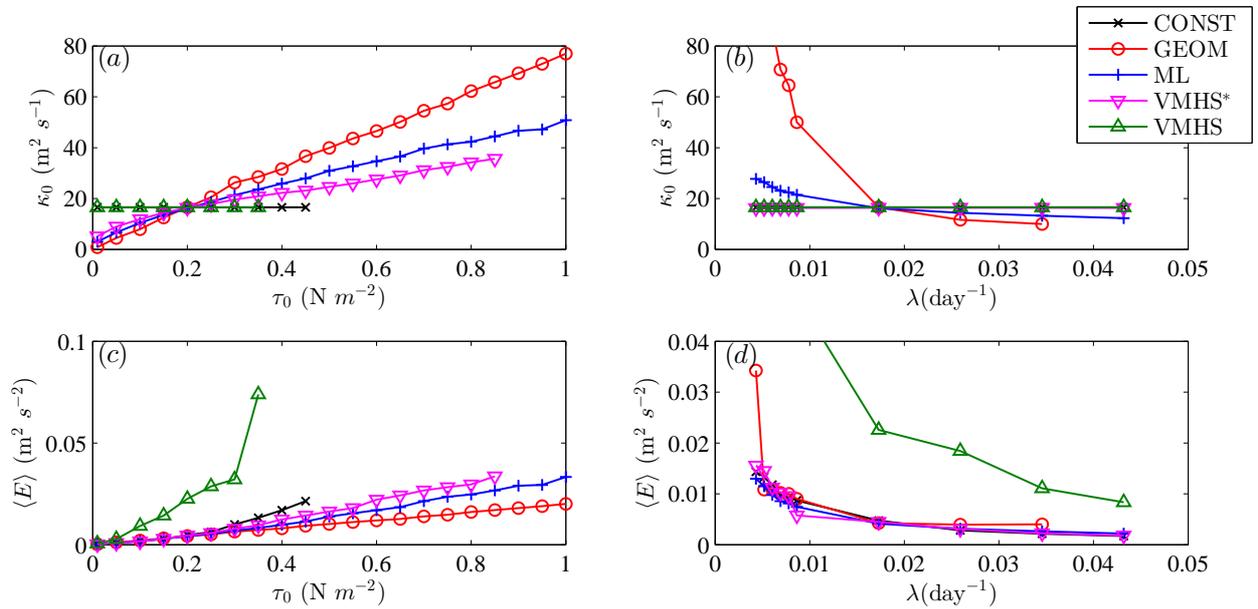}
  \caption{Emergent $\kappa_0$ (panel $a,b$, normalised by
  $N^2_{\textnormal{ref}} = 5\times 10^{-6}$) and domain integrated eddy energy
  $\langle E\rangle$ (panels $c,d$) at varying peak wind forcing $\tau_0$
  (panels $a,c$) and eddy energy dissipation rate $\lambda$ (panels $b,d$), for
  the five parameterisation variants. Note that the value of $\kappa_0$ plays no
  dynamical role for the VMHS variant, and is simply set to the value used for
  the CONST variant. Non-converged solutions have been omitted. Some data points
  are out of the axes range for small $\lambda$ for display reasons.}
  \label{fig:kappa_E_varyparam_space}
\end{figure}

Figure~\ref{fig:kappa_E_varyparam_space} shows the emergent $\kappa_0$ (defined
by taking $N^2_{\textnormal{ref}} = 5\times10^{-6}\ \mathrm{s}^{-1}$) and
$\langle E \rangle$ for varying $\tau_0$ and $\lambda$. The CONST and VMHS
variant has $\kappa_0$ fixed by definition. As before, for varying $\tau_0$, a
near linear trend of $\kappa$ with $\tau_0$ is seen in the GEOM variant, whilst
a sub-linear trend is seen for the ML variant. Varying $\lambda$ again does not
affect CONST, VMHS${}^*$, or VMHS by definition, while this has some effect on
the ML variant and somewhat larger effect on the GEOM variant. For the ML and
especially GEOM variants, increasing $\lambda$ decreases $\kappa_0$. Further,
the eddy energy level is found to decrease with increased dissipation.

\begin{figure}
  \includegraphics[width=\linewidth]{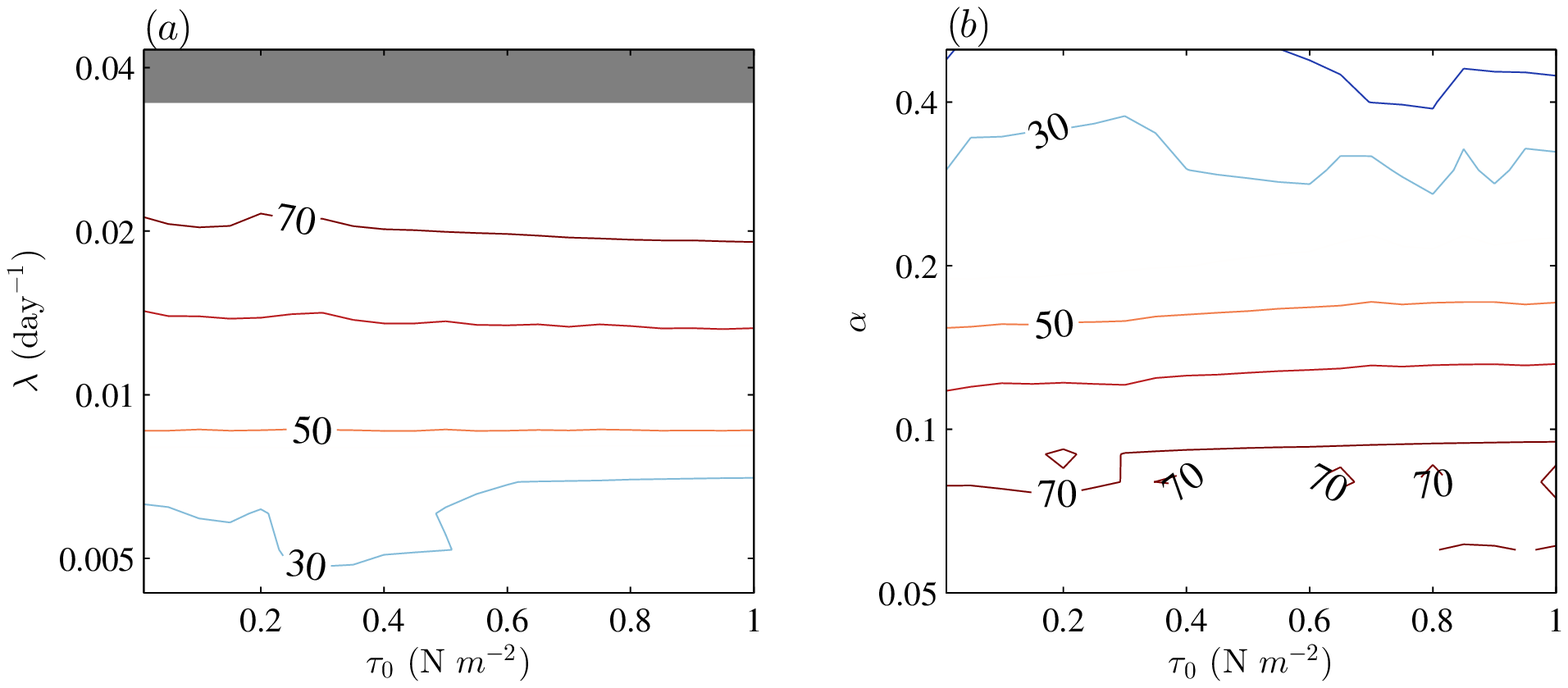}
  \caption{Contour plot of the emergent transport over ($a$) ($\tau_0, \lambda$)
  space (with $\alpha = 0.1$), and ($b$) ($\tau_0, \alpha$) space (with $\lambda
  = 2\times 10^{-7}\ \mathrm{s}^{-1}$). Regions with non-converged solutions
  have been greyed out and their output set to NaN.}
  \label{fig:transport_contour_space}
\end{figure}

\begin{figure}
  \includegraphics[width=\linewidth]{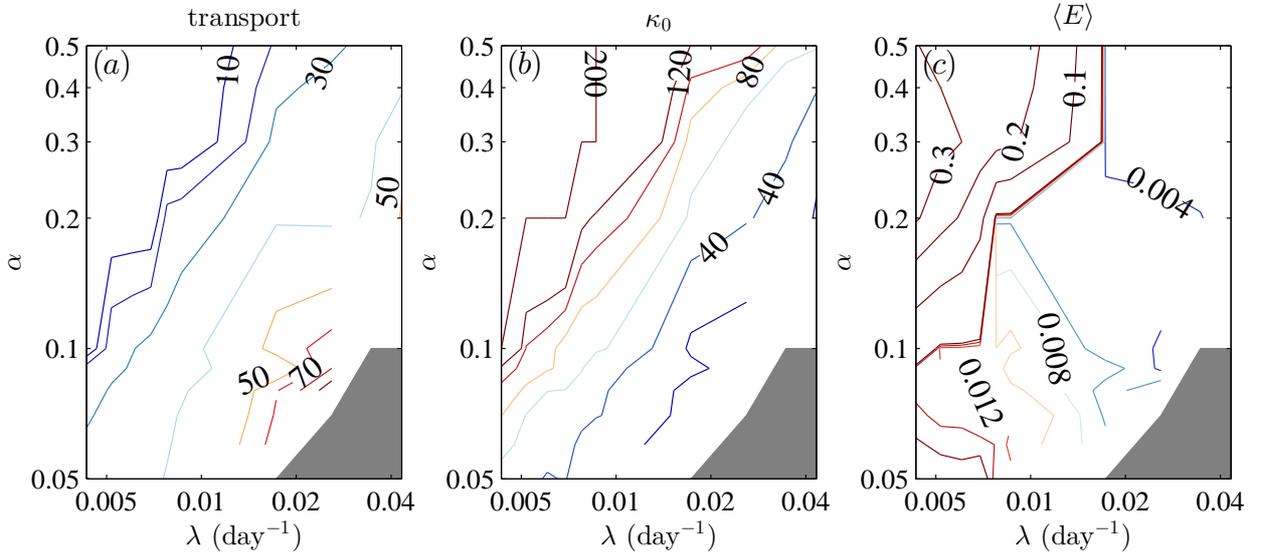}
  \caption{Contour plot of emergent ($a$) transport (in Sv), ($b$) $\kappa_0$
  (normalised by $N^2_{\textnormal{ref}} = 5\times 10^{-6}\ \mathrm{s}^{-1}$, in
  $\ \mathrm{m}^2\ \mathrm{s}^{-1}$) and ($c$) $\langle E \rangle$ (in $\
  \mathrm{m}^2\ \mathrm{s}^{-2}$) of the GEOM variant over $\lambda$ and
  $\alpha$ space, at $\tau_0 = 0.2\ \mathrm{N}\ \mathrm{m}^{-2}$. The axes are
  logarithmic with simulation data from 10 values of $\alpha$ ranging from
  $0.05$ to $0.5$ and 10 values of $\lambda$ ranging from $5\times 10^{-8}\
  \mathrm{s}^{-1}$ to $5\times 10^{-7}\ \mathrm{s}^{-1}$. Regions with
  non-converged solutions have been greyed out and their output set to NaN.}
  \label{fig:emergent_contour_space}
\end{figure}

The emergent eddy saturation for the GEOM variant is again found to be robust
over a range of parameters, as shown in
Figure~\ref{fig:transport_contour_space}, though there is an increased
variability with varying peak wind stress value $\tau_0$ at the smaller
transports. Figure~\ref{fig:emergent_contour_space} shows contour plots of the
emergent properties with varying $\lambda$ and $\alpha$. In both figures,
non-converged states have been greyed out. Although showing much more
variability than the analogous spatially constant $\kappa$ case in
Figure~\ref{fig:emergent_contour}, there is a pattern of increased transport at
increasing $\lambda$ or decreasing $\alpha$, and of decreased $\kappa_0$ at
increasing $\lambda$ or decreasing $\alpha$. Note the region with low $\lambda$
and large $\alpha$ with small transport, large $\langle E\rangle$ and thus large
$\kappa_0$. The resulting parameterised eddies in this case are very strong,
resulting in essentially flat isopycnals.

\section{Conclusions}\label{sec:conclusion}


\subsection{Summary of results}

In this article the problem of emergent eddy saturation in coarse resolution
ocean modelling with parameterised mesoscale eddies has been considered.
Specifically, an idealised zonally averaged channel configuration was used to
test the sensitivity of mean zonal transports with respect to the strength of
surface wind forcing, and additionally with respect to the strength of total
eddy energy dissipation and closure parameters. Variants of the
\cite{GentMcWilliams90} scheme have been tested, with a constant GM coefficient,
a GM coefficient with a stratification dependence based upon that described in
\cite{Visbeck-et-al97}, a GM coefficient with a mixing-length inspired energy
dependence \citep[e.g.,][]{EdenGreatbatch08}, and a GM coefficient derived from
the geometric framework described by \cite{Marshall-et-al12}. For the schemes
with eddy energy dependence a parameterised equation for the domain integrated
eddy energy was solved. By integrating over the domain, specific closures were
derived, falling into two classes --- one where the GM coefficient was spatially
constant, and one where the GM coefficient had a spatial structure based upon
that described in \cite{Ferreira-et-al05}. A form with additional stratification
dependence, closer to the original proposal of \cite{Visbeck-et-al97}, was
additionally tested.

It was found that the scheme derived from the geometric framework of
\cite{Marshall-et-al12} led to almost complete emergent eddy saturation, with
little or no significant dependence of the mean transport on the surface wind
stress magnitude. This was additionally observed for a wide range of eddy energy
dissipation time scales and parameterisation parameter values. Moreover, it was
found that the changes to the equilibrium stratification profile with different
values of peak wind stress were small (not shown). This was found both for the
case where the GM coefficient was spatially constant, and where the GM
coefficient was assumed to depend upon the vertical stratification. Furthermore,
the dependence of the transport and other emergent quantities are consistent
with the physical and mathematical arguments given in \S\ref{sec:mmb}. On the
other hand, the use of a basic spatially and temporally constant GM coefficient
led to a very significant dependence of the mean zonal transport with respect to
the wind stress, similar to behaviour reported in low resolution ocean model
tests described in \cite{Munday-et-al13}. Variants based upon the
\cite{Visbeck-et-al97} and upon mixing length arguments were generally found to
have a somewhat reduced sensitivity, but did not exhibit full eddy saturation.


\subsection{Discussion and future work}

This work focuses on eddy saturation, but an equally important process that has
not been investigated in this work is the ability of the GM coefficient variants
in showing eddy compensation. In particular, the extent of eddy compensation
depends upon both the magnitude and the location of the eddy induced transport,
and the degree to which it cancels with the local Eulerian circulation
\citep{Meredith-et-al12}. The model considered in this article has no
representation for ocean basins and hence is unsuitable for studying eddy
compensation. An investigation into the ability of the \cite{Marshall-et-al12}
variant of the GM coefficient in showing emergent eddy compensation would
require a more sophisticated eddy energy budgets than the one employed here, and
is left as future work.

Assuming that the eddy energy is given via a parameterised eddy energy budget,
the only remaining freedom in the \cite{Marshall-et-al12} variant is in the
specification of the non-dimensional geometric parameter $\alpha$, as all
dimensional information on the magnitude of the GM coefficient is already
provided by the eddy energy and mean stratification. In this work $\alpha$ was
chosen to have a constant value of $0.1$, which was guided by the diagnoses of
the equilibrated states in a three-layer wind forced quasi-geostrophic double
gyre simulation \citep{Marshall-et-al12}, an Eady spindown simulation of the
hydrostatic primitive equations \citep{Bachman-et-al16} as well as a
Phillips-like quasi-geostrophic baroclinic jet spindown problem (Mak \emph{et
al.,}, in preparation). In diagnostic calculations $\alpha$ is not a constant,
and in particular $\alpha$ was found in \cite{Bachman-et-al16} to vary depending
on whether the system is in a linear growth phase or in later phases of the
spindown evolution. It is perhaps of theoretical interest to have $\alpha$
evolving in time to capture the initial instability, finite-amplitude regime,
and transition into an equilibrated state, although this is beyond the scope of
the current work.

In this paper we have found that the functional dependence for the GM
coefficient proposed in \cite{Marshall-et-al12}, which incorporates energetic
constraints through the solution of a parameterised eddy energy budget, yields
near total emergent eddy saturation in a highly idealised configuration. For
implementation in a global ocean model, since the GM scheme is normally built
into the model architecture, it would appear the main additional challenge would
be (i) to add a parameterised eddy energy budget that couples with the GM
scheme, and (ii) derive an appropriate form for a local parameterised eddy
energy budget. The domain integrated eddy energy budget employed here is much
too restrictive in a global ocean model. With this, we envisage the scheme may
be implemented into an operational global circulation model as follows:
\begin{enumerate}
  \item Solve for the provisional eddy transport velocities, with a preferred
  vertical profile for the eddy transfer coefficient, utilising the standard GM
  scheme;
  \item Vertically integrate the implied eddy form stress and compare with the
  theoretical prediction derived from the \cite{Marshall-et-al12} geometric
  framework, with a prescribed non-dimensional parameter $\alpha$;
  \item Solve for the parameterised, vertically integrated eddy energy budget,
  analogous to \cite{EdenGreatbatch08} but for the full, rather than kinetic,
  eddy energy;
  \item Rescale the eddy transport velocities, equivalent to rescaling the GM
  eddy transfer coefficient, uniformly over the vertical column such that the
  vertical integral of the eddy form stress matches the theoretical prediction
  from the \cite{Marshall-et-al12} geometric framework.
\end{enumerate}
By applying the energetic constraint in the vertical integral of the eddy form
stresses, the recipe given above succeeds in retaining the positive-definite
conversion of mean to eddy energy associated with the GM scheme, as well as the
derived energetic constraint given in the \cite{Marshall-et-al12} geometric
framework.

In a closure for ocean turbulence one must typically tune the closure parameters
in order to match a desired large-scale or mean state of interest. However for
many key questions in physical oceanography, it is not only the mean state
itself, but also the sensitivity of that mean state to external changes, which
is of interest. This is, for example, critical to the understanding of the long
time response of the ocean and broader climate system to long term forcing
changes. The Gent--McWilliams closure is now a key component in large scale
climate relevant ocean modelling, but it has been found that existing variants
of the scheme in wide use, in particular with a constant Gent--McWilliams
coefficient, do not yield accurate representations of ocean transport
sensitivities with respect to changed in wind forcing
\citep[e.g.,][]{FarnetiGent11, GentDanabasoglu11}. This work provides the first
evidence that the phenomenon of eddy saturation may be captured without major
changes to the existing Gent--McWilliams closure, simply by employing the
\cite{Marshall-et-al12} form for the GM coefficient, derived from first
principles with no tunable dimensional parmaeters, coupled with a parameterised
eddy energy budget. A proposal on how this scheme may be implemented into a
global circulation model via the addition of a parameterised eddy energy
equation has been given here. Investigations into implementing this into a
general circulation model, as well as theoretical developments for a
parameterised eddy energy budget, are under investigation.


\section*{Acknowledgements}

This work was funded by the UK Natural Environment Research Council grant
NE/L005166/1. The data used for generating the plots in this article is
available through the Edinburgh DataShare service at
\verb!http://dx.doi.org/10.7488/ds/1481!. JM thank Jonas Nycander for
discussions relating to relation \eqref{eq:slope-constraint}. JRM and JM thanks
Malte Jansen and Maarten Ambaum for useful discussions.


\appendix

\section{Eddy energetics}\label{app:eddy_energy}

In \S\ref{sec:gm} the integrated mean energy equation is considered. Here
a corresponding integrated eddy energy equation is derived.

Eddy equations, associated with the mean equations \eqref{eqn:mean_momentum}, are
\begin{linenomath*}\begin{subequations}
  \begin{align}
    & \frac{\partial u'^z}{\partial t}
      + u \frac{\partial u'^z}{\partial x}
      + v \frac{\partial u'^z}{\partial y}
      + w \frac{\partial u'^z}{\partial z}
      + u'^z \frac{\partial \overline{u}^z}{\partial x}
      + v'^z \frac{\partial \overline{u}^z}{\partial y}
      + w'^z \frac{\partial \overline{u}^z}{\partial z}
      - f v'^z \nonumber \\
    & \qquad ~ =
      - \overline{u'^z \frac{\partial u'^z}{\partial x}}^z
      - \overline{v'^z \frac{\partial u'^z}{\partial y}}^z
      - \overline{w'^z \frac{\partial u'^z}{\partial z}}^z    
      - \frac{1}{\rho_0} \frac{\partial p'^z}{\partial x} + F'^z - D'^z, \label{aeqn:eddy_mom_u} \\
    & \frac{\partial v'^z}{\partial t}
      + u \frac{\partial v'^z}{\partial x}
      + v \frac{\partial v'^z}{\partial y}
      + w \frac{\partial v'^z}{\partial z}
      + u'^z \frac{\partial \overline{v}^z}{\partial x}
      + v'^z \frac{\partial \overline{v}^z}{\partial y}
      + w'^z \frac{\partial \overline{v}^z}{\partial z}
      + f u'^z \nonumber \\
    & \qquad ~ =
      - \overline{u'^z \frac{\partial v'^z}{\partial x}}^z
      - \overline{v'^z \frac{\partial v'^z}{\partial y}}^z
      - \overline{w'^z \frac{\partial v'^z}{\partial z}}^z    
      - \frac{1}{\rho_0} \frac{\partial p'^z}{\partial y}, \label{aeqn:eddy_mom_v} \\
    & \qquad 0 = -\frac{1}{\rho_0} \frac{\partial p'^z}{\partial z} - \frac{g \rho'^z}{\rho_0}. \label{aeqn:eddy_hydrostatic}
  \end{align}
\end{subequations}\end{linenomath*}
$\left( \ldots \right)'^z$ denotes an eddy component associated with a zonal
average at fixed height -- for example $\rho'^z = \rho - \overline{\rho}^z$. It
is assumed throughout this section that $f'^z = 0$, and that $g$ and $\rho_0$
are spatially and temporally constant. In the following it is further assumed
that the mean and eddy velocities $\left( 0, \overline{v}^z, \overline{w}^z
\right)^T$ and $\left( u'^z, v'^z, w'^z \right)^T$ are incompressible and have
zero normal component on domain boundaries.

Multiplying equation \eqref{aeqn:eddy_mom_u} by $u'^z$, equation
\eqref{aeqn:eddy_mom_v} by $v'^z$, zonally averaging at constant height, using
the hydrostatic relation \eqref{aeqn:eddy_hydrostatic}, and integrating over the
domain, leads to the integrated eddy kinetic energy budget
\begin{linenomath*}\begin{align}\label{aeqn:eke}
  \iint \rho_0 K \mathrm{d}y\, \mathrm{d}z = & \iint \rho_0 \overline{u'^z \left( F'^z - D'^z \right)}^z \mathrm{d}y\, \mathrm{d}z \nonumber \nonumber \\
  - & \iint \rho_0 \left[ \frac{\partial \overline{u}^z}{\partial y} \overline{u'^z v'^z}^z + \frac{\partial \overline{u}^z}{\partial z} \overline{u'^z w'^z}^z + \frac{\partial \overline{v}^z}{\partial y} \overline{v'^z v'^z}^z + \frac{\partial \overline{v}^z}{\partial z} \overline{v'^z w'^z}^z \right] \mathrm{d}y\, \mathrm{d}z, \nonumber \\
  - & \iint g \overline{w'^z \rho'^z}^z \mathrm{d}y\, \mathrm{d}z,
\end{align}\end{linenomath*}
with eddy kinetic energy (per unit volume)
\begin{linenomath*}\begin{equation}
  \rho_0 K = \frac{1}{2} \rho_0 \overline{u'^z u'^z}^z + \frac{1}{2} \rho_0 \overline{v'^z v'^z}^z.
\end{equation}\end{linenomath*}
Now from the density equation
\begin{linenomath*}\begin{equation}
  \frac{\partial \rho}{\partial t} + \frac{\partial \left( u \rho \right)}{\partial x} + \frac{\partial \left( v \rho \right)}{\partial y} + \frac{\partial \left( w \rho \right)}{\partial z} = 0,
\end{equation}\end{linenomath*}
multiplying by the height $z$, zonally averaging at constant height, and
integrating over the domain, leads to
\begin{linenomath*}\begin{align}\label{aeqn:eddy_w_density}
  \iint \overline{w'^z \rho'^z}^z \mathrm{d}y\, \mathrm{d}z
    & = \iint \frac{\partial \left( \overline{\rho}^z z \right)}{\partial t} \mathrm{d}y\, \mathrm{d}z - \iint \overline{w}^z \overline{\rho}^z \mathrm{d}y\, \mathrm{d}z \nonumber \\
    & = \iint \frac{\partial \left( \overline{\rho}^\rho z \right)}{\partial t} \mathrm{d}y\, \mathrm{d}z - \frac{\partial \left( \overline{\rho}^* z \right)}{\partial t} \mathrm{d}y\, \mathrm{d}z  - \iint \overline{w}^z \overline{\rho}^z \mathrm{d}y\, \mathrm{d}z,
\end{align}\end{linenomath*}
where $\rho^* = \overline{\rho}^\rho - \overline{\rho}^z$ is the difference
between the two mean densities. Assuming that the eddy transport velocity
$\left( 0, v^*, w^* \right)^T$ has zero normal component on domain boundaries,
multiplying equation \eqref{eqn:mean_density} by the height $z$ and integrating
over the domain yields
\begin{linenomath*}\begin{equation}\label{aeqn:mean_pe}
  \iint \frac{\partial \left( \overline{\rho}^\rho z \right)}{\partial t} \mathrm{d}y\, \mathrm{d}z
    = \iint \hat{w} \overline{\rho}^\rho \mathrm{d}y\, \mathrm{d}z.
\end{equation}\end{linenomath*}
Combining equations \eqref{aeqn:eke}, \eqref{aeqn:eddy_w_density}, and
\eqref{aeqn:mean_pe} leads to the integrated eddy energy budget
\begin{linenomath*}\begin{align}\label{aeqn:integrated_ee}
  \frac{\mathrm{d}}{\mathrm{d} t} \iint \rho_0 E \mathrm{d}y\, \mathrm{d}z
  & = \iint \rho_0 \overline{u'^z \left( F'^z - D'^z \right)}^z \mathrm{d}y\, \mathrm{d}z \nonumber \\
  & - \iint \rho_0 \left[ \frac{\partial \overline{u}^z}{\partial y} \overline{u'^z v'^z}^z + \frac{\partial \overline{u}^z}{\partial z} \overline{u'^z w'^z}^z + \frac{\partial \overline{v}^z}{\partial y} \overline{v'^z v'^z}^z + \frac{\partial \overline{v}^z}{\partial z} \overline{v'^z w'^z}^z \right] \mathrm{d}y\, \mathrm{d}z \nonumber \\
  & - \iint w^* g \overline{\rho}^\rho \mathrm{d}y\, \mathrm{d}z - \iint \overline{w}^z g \rho^* \mathrm{d}y\, \mathrm{d}z,
\end{align}\end{linenomath*}
with total eddy energy (per unit volume)
\begin{linenomath*}\begin{equation}
  \rho_0 E = \frac{1}{2} \rho_0 \overline{u'^z u'^z}^z + \frac{1}{2} \rho_0 \overline{v'^z v'^z}^z - \rho^* g z.
\end{equation}\end{linenomath*}

The first right-hand-side term in equation \eqref{aeqn:integrated_ee} is the
eddy energy generation due to forcing in the horizontal momentum equations. The
second right-hand-side term is the mean-to-eddy energy conversion due to the
eddy Reynolds stresses. The third right-hand-side term is the mean-to-eddy
energy generation due to the eddy transport velocity, and corresponds exactly to
the conversion term appearing in the mean energy equation
\eqref{eqn:mean_energy}. The final term is an additional conversion term which
arises from the direct application of an average at constant height to the
hydrostatic relation \citep[see the discussion in][appendix
B]{McDougallMcIntosh01}. Replacing $\overline{\rho}^\rho$ with
$\overline{\rho}^z$ in equation \eqref{eqn:mean_hydrostatic} would lead to the
appearance of a corresponding term in the integrated mean energy equation.


\section{Deriving equation \eqref{eq:slope-constraint}}\label{app:details}

If both the GM coefficient and the eddy energy dissipation scale with the eddy
energy, then there is an apparent degeneracy in the eddy energy equation
\eqref{eqn:eddy_energy}. If, for example, the scaling factors are
constant, then the integrated eddy energy can be factored out, leading to a
balance between the rates of eddy energy generation and dissipation. In this
appendix this property is formalised somewhat via the use of appropriate
integral inequalities.

It is assumed that functions $f$ and
$g$ are suitably smooth such that H\"older's inequality \cite[e.g.,][Appendix
A]{DoeringGibbon-Analysis}
\begin{linenomath*}\begin{equation*}
  \left\|fg\right\|_{L^1} \leq \left\|f\right\|_{L^p} \left\|g\right\|_{L^q},
  \qquad \left\|f\right\|_{L^p} = 
    \left(\int_\Omega |f|^p\, \mathrm{d}\Omega\right)^{1/p},
  \qquad \frac{1}{p} + \frac{1}{q} = 1,
\end{equation*}\end{linenomath*}
may be applied. Choosing the H\"older conjugates $p = 2$ and $q = 2$ (i.e. a
generalised Cauchy--Schwartz inequality) and applying the above inequality to
the steady state eddy energy equation \eqref{eqn:eddy_energy} leads to
\begin{linenomath*}\begin{equation}
  \lambda\left\|E\right\|_{L^1} \leq 
    \alpha\left\|E\right\|_{L^2}
    \left\| \frac{|f|\rho_0}{gN} \left|\ddy{\overline{u}^z}{z}\right|\right\|_{L^2}.
\end{equation}\end{linenomath*}
Notice that the $L^1$ norm is the integral of the absolute value, and so
$\left\|E\right\|_{L^1} = \iint E\, \mathrm{d}y\, \mathrm{d}z$. From this, it
follows that
\begin{linenomath*}\begin{equation}
  \frac{\lambda}{\alpha}\frac{\|E\|_{L^1}}{\|E\|_{L^2}} = 
    \tilde{C} \left\| \frac{|f|\rho_0}{gN} \left|\ddy{\overline{u}^z}{z}\right|\right\|_{L^2},
\end{equation}\end{linenomath*}
for $\tilde{C}\in(0,1]$. Although $\|E\|_{L^1}\leq\|E\|_{L^2}$ (a consequence of
H\"older's inequality), if $\|E\|_{L^1}\approx\|E\|_{L^2}$ then the relation
\eqref{eq:slope-constraint} results.

Some more progress may be made if the norms of the derivatives may be assumed to
be small. Assuming a bounded Lipschitz domain, the $\left\|E\right\|_{L^2}$ term
may be controlled by utilising the Gagliardo--Nirenberg interpolation inequality
(\citealt{Nirenberg59}, \S2; see also Appendix A of
\citealt{DoeringGibbon-Analysis}), which states that
\begin{linenomath*}\begin{equation*}
  \|\mathrm{D}^j f\|_{L^p} \leq C_1 \|\mathrm{D}^m f\|_{L^r}^a \|f\|_{L^q}^{1-a} + C_2 \|f\|_{L^s}, \qquad \frac{1}{p} = \frac{j}{d} + \left( \frac{1}{r} - \frac{m}{d}\right) a + \frac{1-a}{q}
\end{equation*}\end{linenomath*}
with $\mathrm{D}$ being a weak derivative, $d$ is the dimensionality of the
domain, $1\leq r,q\leq\infty$, $j/m\leq a\leq 1$, and $s>0$ is arbitrary. This
does not cover some exceptional cases, though they are not of interest here. The
constants $C_{1,2}$ only depend on the domain and the choice of the parameter
values. For $d=2$ here, taking $j=0$, $p=2$ and $s=1$, it is noted that $m=1$,
$r=1$ and $a=1$ is one option (which is a form of the Sobolev inequality; e.g.,
\citealt[][\S5.6.1]{Evans-PDE}), and taking $s=1$ and $f=E$ results in
\begin{linenomath*}\begin{equation*}
  \|E\|_{L^2} \leq C_1 \|\mathrm{D}E\|_{L^1} + C_2 \|E\|_{L^1}.
\end{equation*}\end{linenomath*}
If $m=1$, $r=2$, $a=1/2$ and $q=1$ instead, then
\begin{linenomath*}\begin{equation*}
  \|E\|_{L^2} \leq C_1 \|\mathrm{D}E\|_{L^2}^{1/2} \|E\|_{L^1}^{1/2} + C_2 \|E\|_{L^1},
\end{equation*}\end{linenomath*}
which is analogous to the inequality of \cite{Nash58}. Other possibilities exist
involving higher derivatives. Either way, assuming that the terms involving the
derivatives are small compared to $C_2\|E\|_{L^1}$, then the relation
\eqref{eq:slope-constraint} again follows, with a constant of proportionality
that only depends on the domain and the parameter values chosen in the
Gagliardo--Nirenberg interpolation inequality and is bounded away from zero and
infinity.

The Gagliardo--Nirenberg inequality may be applied again to further control the
term involving the derivative in terms of higher $L^p$ norms, although the
relation again needs further assumptions. H\"older's inequality with the
conjugates $p = 1$ and $q=\infty$ could be applied at the beginning to factor
out $\|E\|_{L^1}$ immediately, however there is then a lack of control on the
$L^\infty$ norm as the Gagliardo--Nirenberg inequality above does not apply. An
alternative approach may be to consider adding and subtracting the mean of the
relevant function and apply the Minkowski inequality (a generalised triangle
inequality) which would yield similar results. This approach when applied to the
ML variant $\kappa = \alpha_{\tiny \mbox{ML}}$ does not appear to yield the same
bound as the eddy energy equation does not become degenerate.




\bibliographystyle{elsarticle-harv} 
\bibliography{refs}


\end{document}